\documentclass[12pt]{article}

\usepackage{preprint}

\usepackage{amsmath, amsthm, amssymb, amsfonts}

\usepackage[numbers,square]{natbib}
\usepackage[utf8]{inputenc}
\usepackage[T1]{fontenc}
\usepackage{xcolor}
\usepackage[colorlinks=true,
            linkcolor=purple,
            urlcolor=blue,
            citecolor=cyan,
            anchorcolor=black]{hyperref}
\usepackage{booktabs}
\usepackage{longtable}
\usepackage{nicefrac}
\usepackage{microtype}
\usepackage{float}
\usepackage{graphicx}
\usepackage{multirow}
\usepackage{array}
\providecommand{\arraybackslash}{\let\\=\tabularnewline}
\usepackage{algorithmic}
\usepackage{algorithm}

\usepackage{setspace}
\usepackage{newfloat}
\DeclareFloatingEnvironment[name={Supplementary Figure}]{suppfigure}
\usepackage{sidecap}
\sidecaptionvpos{figure}{c}

\usepackage{titlesec}
\titleformat{\section}[block]{\large\bfseries\raggedright}{}{0em}{\MakeUppercase}
\titleformat{\subsection}[block]{\normalsize\bfseries\raggedright}{}{0em}{}
\titleformat{\subsubsection}[block]{\normalsize\mdseries\raggedright}{}{0em}{}
\titleformat{\paragraph}[runin]{\normalsize\itshape}{}{0em}{}
\titlespacing\section{0pt}{14pt plus 4pt minus 3pt}{4pt plus 2pt minus 1pt}
\titlespacing\subsection{0pt}{12pt plus 3pt minus 3pt}{2pt plus 1pt minus 1pt}
\titlespacing\subsubsection{0pt}{10pt plus 3pt minus 3pt}{1pt plus 1pt minus 1pt}

\usepackage{tikz}

\title{Documenting Trofinetide Treatment Phenotypes in Rett Syndrome with a
Dual-LLM Consensus Pipeline}
\date{}

\SetBgContents{}

\begin{document}

\thispagestyle{empty}

\begin{center}
{\LARGE\bfseries Documenting Trofinetide Treatment Phenotypes in Rett
Syndrome with a Dual-LLM Consensus Pipeline\par}
\vspace{0.4cm}
{\small Article type: Research and Applications}
\end{center}

\vspace{0.4cm}

\noindent\textbf{Authors.}
Daniel Palacios, BS\textsuperscript{1,2,3,4},
Guantong Qi, BS\textsuperscript{2,3,4,5},
Mikael Guzman~Karlsson, MD, PhD\textsuperscript{2,6},
Shawntel K. David, BSN, RN\textsuperscript{7},
Mohammed Sameeruddin, BSN, RN\textsuperscript{7},
Bernhard Suter, MD\textsuperscript{2,7,8},
Zhandong Liu, PhD\textsuperscript{1,2,3,4,*},
Davut Pehlivan, MD\textsuperscript{2,3,7,8,*},
Hu Chen, PhD\textsuperscript{2,3,4,*}.

\smallskip
\noindent\textsuperscript{*}Co-corresponding authors.

\medskip
\noindent\textbf{Affiliations.}
\begin{description}\setlength{\itemsep}{0pt}
\item[1.] Quantitative and Computational Biosciences, Baylor College of Medicine, Houston, Texas, 77030, USA
\item[2.] Department of Pediatrics, Baylor College of Medicine, Houston, Texas, 77030, USA
\item[3.] Jan and Dan Duncan Neurological Research Institute, Texas Children's Hospital, Houston, Texas, 77030, USA
\item[4.] Data Science Center, Texas Children's Hospital, Houston, Texas, 77030, USA
\item[5.] Genetics \& Genomics Graduate Program, Baylor College of Medicine, Houston, Texas, 77030, USA
\item[6.] Department of Pathology and Immunology, Baylor College of Medicine, Houston, Texas, 77030, USA
\item[7.] Blue Bird Circle Rett Center, Texas Children's Hospital, Houston, Texas, 77030, USA
\item[8.] Section of Pediatric Neurology and Developmental Neuroscience, Department of Pediatrics, Baylor College of Medicine, Houston, Texas, 77030, USA
\end{description}

\medskip
\noindent\textbf{Corresponding author.}
Hu Chen, Department of Pediatrics, Baylor College of Medicine, and Jan and Dan
Duncan Neurological Research Institute, Texas Children's Hospital, 1250 Moursund
Street, Houston, TX 77030, USA. Telephone: 832-652-0576.
Email: \texttt{Hu.Chen@bcm.edu}.

\medskip
\noindent\textbf{Keywords:}
Rett syndrome; trofinetide; large language models; natural language processing;
electronic health records.

\medskip
\noindent\textbf{Word count.} Abstract: 247. Main body: 3,859.
Tables: 4. Figures: 3.
\clearpage

\begin{center}\large\bfseries ABSTRACT\end{center}

\noindent\textbf{Objective.}
Trofinetide, approved in 2023 as the first Rett syndrome treatment, lacks
scalable real-world monitoring because chart review is labor-intensive. We
developed a dual large language model (LLM) consensus pipeline to extract 18
treatment-response phenotypes, comprising 4 adverse effects, 13 Rett
clinical features, and medication refusal, from clinical notes.

\smallskip
\noindent\textbf{Materials and Methods.}
We analyzed 26,763 notes from 108 patients with confirmed sustained
trofinetide exposure at Texas Children's Hospital. The pipeline filters notes by keyword,
classifies each through two LLMs requiring consensus, and aggregates a
patient-by-timepoint matrix. Architecture and prompt were selected on a diarrhea
development set, then frozen and applied to the other 17 phenotypes with the
phenotype definition substituted, as an out-of-sample test. The reported prompt comparison scores 183 expert-labeled
cells.

\smallskip
\noindent\textbf{Results.}
A prompt requiring diarrhea to be active and severe performed
best (84.2$\pm$0.0\% accuracy; F1$=$0.674), 47 percentage points above
regular-expression matching ($p<0.0001$). Expert reviewers judged six of the
pipeline's calls per phenotype, three positive and three negative, and agreed with 90.2\%
(92/102). The 13 Rett clinical
features are manifestations of the disorder itself rather than adverse effects
of trofinetide. Gross and fine motor impairment, language/communication, and
stereotypy were documented present in 92--100\% of patients at
every visit.

\smallskip
\noindent\textbf{Discussion.}
In-context LLM extraction captured multi-domain treatment responses. The
outcome is documented presence at each visit, not a direction-of-change
judgment.

\smallskip
\noindent\textbf{Conclusion.}
Coupling in-context LLM reasoning with a consensus architecture offers a
scalable alternative to keyword and regular-expression extraction for real-world evidence,
adapting to new phenotypes and medications through configuration alone.

\clearpage

\section*{Introduction}
\label{sec:introduction}

Rett syndrome is a severe neurodevelopmental disorder affecting approximately
1 in 10,000--15,000 live female births, caused primarily by loss-of-function
mutations in the \textit{MECP2} gene~\cite{amir1999rett,neul2010rett}.
The condition is characterized by an initial period of apparently normal
development followed by a period of regression, with loss of acquired hand
skills, development of stereotypic hand movements, gait abnormalities, and
loss of spoken language~\cite{neul2010rett}, followed by a period of
stabilization without further regression of skills.

In March 2023, trofinetide (Daybue) became the first FDA-approved treatment
for Rett syndrome~\cite{fda2023daybue}, following earlier
phase~2 efficacy data~\cite{glaze2017double} and a pivotal Phase~3 trial
demonstrating improvements on the Rett Syndrome Behaviour Questionnaire
(RSBQ) and Clinical Global Impression--Improvement (CGI-I)
scale~\cite{neul2023trofinetide}. However, the trial captured only a subset
of treatment-relevant phenotypes over 12 weeks, and real-world treatment
response across the full spectrum of Rett syndrome features remains
incompletely characterized.

\section*{Background and Significance}
\label{sec:background}

Monitoring treatment response in Rett syndrome is challenging. The condition
manifests across multiple domains, including motor, cognitive, behavioral and
autonomic function, each potentially documented by a different specialist in
narrative notes. Manual chart review of the resulting note volume can
quickly become prohibitively labor-intensive. Electronic health record (EHR) phenotyping has therefore
turned to automated methods,
from concept-extraction systems~\cite{savova2010mayo,fu2020clinical} to
rule-based and portable phenotype algorithms~\cite{shivade2014review,kirby2016phekb}.
But traditional rule-based natural language processing
(NLP)~\cite{wang2018clinical,kreimeyer2017natural} struggles to distinguish
active symptoms from historical, anticipatory, and negated mentions within
complex narratives~\cite{chapman2001negex,harkema2009context}, a difficulty
long noted in adverse-event detection~\cite{murff2011automated,harpaz2012novel}.

Large language models (LLMs) perform strongly on clinical text and medical
benchmarks~\cite{singhal2023large,thirunavukarasu2023large,nori2023capabilities,liu2023deid},
offering the contextual reasoning to interpret narratives, but a single LLM
applied directly to clinical extraction can perform poorly~\cite{tang2023does}.
In our data, a single model produced both false positives, over-interpreting
routine documentation, and false negatives. The appropriate architecture
depends on task complexity. Simple extraction tasks may suit a single
well-prompted model~\cite{wang2024rethinkingboundsllmreasoning,cemri2025multiagentllmsystemsfail}.
Our task spans many phenotypes and requires judging, within each note, whether a
finding is active at that visit rather than historical, negated or anticipatory, so it
is complex enough to benefit from a multi-agent verification
design in which one model evaluates or critiques another's
output~\cite{gu2025surveyllmasajudge,mcaleese2024critics}. We therefore use two
independently prompted models and require agreement before making a
positive call, prioritizing precision over recall.

We present a dual-LLM consensus pipeline that extracts 18 treatment-response
phenotypes from clinical notes of 108 patients with Rett syndrome receiving
trofinetide.
The 18 comprise 4 adverse effects, 13 Rett clinical features and medication
refusal, and each is classified at every visit as documented present or absent
(Methods). The
pipeline was developed and evaluated on an expert-annotated ground truth for
the diarrhea phenotype, then applied to the remaining 17 phenotypes with only the
phenotype definition substituted.
Our objective was to
determine whether contextual LLM reasoning, combined with a consensus
requirement, can extract multi-domain treatment-response phenotypes from
routine clinical documentation at a scale that manual review cannot reach. We
focused on one major adverse effect, diarrhea, to develop the pipeline, and we
report that process in full.

\section*{Materials and Methods}
\label{sec:methods}

\subsection*{Study design and population}
\label{sec:study_design}

We conducted a retrospective observational study of patients with Rett syndrome
identified from Rett Center clinical visits at Texas Children's Hospital (TCH)
who were prescribed or considered for trofinetide (Daybue). We selected this
population because the TCH Rett Center is a dedicated center of excellence whose
standardized note templates capture comprehensive disorder-specific information
over office visits of at least one hour.
An initial 115 patients were identified on this basis. To verify sustained
drug exposure,
an independent quality-control (QC) review triangulated three sources per
patient: pharmacy orders, medication-administration records, and longitudinal
note mentions. This confirmed sustained exposure in \textbf{108 of 115 patients
(94\%)}, with 65 by pharmacy order and 43 by clinical documentation alone
($\geq$2 note mentions over $>$60 days). The remaining 7 were unconfirmed (4
with only isolated mentions, 3 with no evidence) and were therefore excluded.
\textbf{This 108-patient QC-confirmed cohort is the analysis cohort throughout}.

Cohort characteristics for the 108 QC-confirmed patients and the 56
diarrhea-validation patients are summarized in
Table~\ref{tab:cohort_characteristics}. The subset is
comparable to the remaining QC-confirmed patients on age, sex, and follow-up,
but has
more notes per patient (median 171 vs.\ 99, $p=0.07$). The full cohort's
median is 148. The per-patient
note count is right-skewed (range 7--2{,}546), summing to
the 26,763 analyzed notes, and the cohort is young (median age 13~years) and
almost entirely female, consistent with Rett syndrome epidemiology. The rare male
patients with \textit{MECP2}-related disorders can follow a different disease
course, but we retained them because this analysis documents phenotype presence
rather than disease severity.

\begin{table}[H]
\centering
\caption{Cohort characteristics for the full QC-confirmed cohort and the
diarrhea-validation subset. Values are median (IQR) or $n$ (\%). The $p$-column
compares the validation subset against the remaining QC-confirmed patients,
Mann--Whitney $U$ for continuous variables and Fisher exact for sex. Variable
definitions, and the patient count each row rests on, are in Supplementary
Note~S8.}
\label{tab:cohort_characteristics}
\begin{tabular}{lccc}
\toprule
\textbf{Characteristic} & \textbf{Full cohort} & \textbf{Validation subset}
& \textbf{$p$} \\
 & \textbf{($n=108$)} & \textbf{($n=56$)} & \\
\midrule
Age at trofinetide start, years                     & 12.7 (8.2--20.3)   & 12.6 (8.3--18.5)   & 0.72 \\
Female, $n$ (\%)                                    & 106 (98.1)         & 56 (100.0)         & 0.23 \\
Follow-up duration, months                          & 29.4 (20.0--34.1)  & 32.4 (25.3--34.1)  & 0.06 \\
Notes analyzed per patient, median (IQR)            & 148 (49--349)      & 171 (61--397)      & 0.07 \\
Notes analyzed per patient, mean                    & 270                & 271                & \\
Total analyzed notes                                & 26,763             & 15,177             & \\
Total notes retrieved                               & 91,453             & 47,491             & \\
\bottomrule
\end{tabular}
\end{table}

\subsection*{Data sources}
\label{sec:data_sources}

We integrated four data sources: the Rett Center registry, from which the
cohort was identified; the clinical-notes repository, spanning all documented
note types; and two structured Epic EHR medication tables (order and
administered). All four reside as ontology-mapped tables in Texas Children's
Hospital's Palantir Foundry environment. The registry is curated by the Rett
Center clinical team of two pediatric nurses and two pediatric neurologists,
who together performed all clinical annotation and review reported here, and
the notes and medication tables are drawn from the institutional EHR.

The QC review found that 43 of the 108 confirmed patients had \emph{no}
matching row in the Epic order table. Trofinetide is a
specialty-pharmacy product whose dispensing often bypasses the institutional
order table, so structured orders alone would have undercounted these
patients, motivating our notes-inclusive design. Retrieval for the QC-confirmed
patients returned 91,453 notes, of which the 26,763 falling inside the six
evaluation windows constitute the analyzed corpus. Retrieval was not restricted to
the treatment period, so the retrieved notes span November 1998 to August 2026,
while the analyzed notes span July 2021 to August 2026.

The retrieved corpus spans 67 note types (Figure~\ref{fig:pipeline}C; full distribution in
Supplementary Table~S1), the largest being telephone encounters at 22.7\% and
progress notes, followed by a large ``unspecified'' group and letters. Because
telephone encounters
are often the first channel through which adverse effects like diarrhea are
reported, we retain all note types rather than restricting to progress
notes.

\subsection*{Pipeline architecture}
\label{sec:pipeline}

The extraction pipeline follows a five-step architecture
(Figure~\ref{fig:pipeline}):

\paragraph{Step 1: Medication Timeline.}
Every follow-up window is measured from the patient's \emph{exposure anchor},
meaning the identified date on which trofinetide was started. The anchor is taken
from the Epic medication-administration record where available, and otherwise from
the earliest note documenting active dosing, adjudicated by LLM review; no patient
anchored on a dispense record,
because trofinetide fills bypass the institutional tables. Across the 108
patients, 8 anchored on an administration record and 91 on an adjudicated note,
while 9 lack a reliable date and are excluded from the timepoint analysis.

\paragraph{Step 2: Note Bucketing.}
Each note is assigned to a six-month bucket (baseline through 30 months) by
the days between its contact date and the exposure anchor. Notes are
then pre-filtered by phenotype-specific keyword
lists (Supplementary Table~S2), producing 60,537 note--phenotype matches across
the 17 phenotypes scored by the presence prompt, all but 7 of which returned a
valid YES/NO in the structured output from both models. Those 7 were scored negative
(Supplementary Note~S1, which also reports inter-LLM agreement). For diarrhea the filter matches 7,111 of the 91,453 retrieved notes,
and the 2,260 of those falling inside an evaluation window are the corpus every
diarrhea comparison below is scored on.

\paragraph{Step 3: Dual-LLM Classification.}
Each keyword-positive note is independently classified by two LLMs with
complementary perspectives. The first, a clinical NLP expert, judges whether the
phenotype is documented in the note itself, applying explicit rules to separate
findings recorded at that visit from historical, negated, anticipatory, and
template mentions against a written per-phenotype definition. For diarrhea that
definition adds a severity threshold. The second, a domain specialist (gastroenterologist for side
effects, pediatric neurologist for the Rett clinical features), weighs clinical
significance and attribution. Prompt templates, YES/NO criteria, and a contingency table of the two
agents' calls (evidence they work independently) are in Supplementary
Notes~S3 and~S1. Both output a
structured YES/NO call with supporting evidence, and a note is positive only
when both models agree YES. The record also carries the note's contact date, which
is known from EHR metadata rather than extracted from the note text (Supplementary
Note~S3). The evidence spans were spot-checked but not systematically validated. Detection is thus \emph{note-local in time}, so
historical mentions score negative, at the cost of missing purely retrospective
ones (quantified in the error analysis).

The LLMs were accessed within Palantir Foundry (Claude~Opus~4.6, July 2026,
default decoding parameters) with output constrained to the structured
YES/NO schema (Supplementary Note~S3). Because the defaults are not
guaranteed deterministic, we report run-to-run variability across three
independent runs.

\paragraph{Step 4: Aggregation.}
For each patient and timepoint, if \emph{any} note in the bucket receives a
positive consensus call, the patient is marked \emph{Yes} for that phenotype, producing
a patient-by-timepoint-by-phenotype matrix.

\paragraph{Step 5: Evaluation.}
Diarrhea was evaluated against a ground truth from independent manual chart
review by domain experts, who labeled each patient-timepoint cell Yes
(medication-associated diarrhea), No, or Deceased/Not~applicable. This produced
labels for 56 patients across the six timepoints
(Table~\ref{tab:cohort_characteristics}), giving 183 evaluable cells inside
follow-up, of which 47 are labeled Yes. Metrics include
accuracy, precision, recall, F1, Cohen's $\kappa$, Matthews correlation, and
per-timepoint confusion matrices.

\emph{Annotation procedure.} The ground truth was established by the Rett
Center clinical team, who reviewed each patient's notes within a timepoint
window using the same labeling scheme. The same reviewers performed the
multi-phenotype expert spot-check. Annotation followed a consensus-adjudication workflow with no
formal written guideline and no retrospective inter-annotator agreement
statistic, a limitation for gold-standard reliability detailed in
Supplementary Note~S6 (TRIPOD-LLM items 8a--8b)~\cite{gallifant2025tripod}.

\subsection*{Phenotype definitions}
\label{sec:phenotypes}

The 18 phenotypes were selected by the Rett Center clinical team as the adverse
effects and disease-related functional domains most salient to monitor in
routine trofinetide care, rather than an exhaustive enumeration. At each
clinical visit, each phenotype is classified according to whether it is
documented as present in that visit's clinical note, not as a
direction-of-change judgment. They span three
categories.

\paragraph{Adverse effects (4):} Diarrhea, Vomiting, Weight Loss, and Other
Side Effects, each a physiological reaction to trofinetide. The diarrhea
phenotype, defined as medication-associated diarrhea, is the target of the
ground-truth validation and error analysis.

\paragraph{Medication refusal (1):} Refusing Medication is tracked
separately because it is a behavioral response, not a physiological
reaction, and captures documented instances of the patient declining or
resisting the medication.

\paragraph{Rett clinical features (13):} General Mood, Gross Motor, Fine Motor,
Language/Communication, Stereotypy, Attention/Alertness, Sleep,
Breathing/Hyperventilation, Bruxism, Dysautonomia, Aggressive Behavior,
Anxiety, Seizures. These are manifestations of Rett syndrome itself, part of
the natural history of the disorder that clinicians track over the course of
treatment. They are not adverse effects, and no finding in this category should
be read as an effect caused by trofinetide.

Each phenotype is defined by a keyword list (Supplementary Table~S2) and
phenotype-specific LLM prompts encoding what constitutes a true positive.
Generic terms matching routine documentation are excluded (e.g., ``alert''
is dropped from the attention/alertness list to avoid vital-sign notes).

\subsection*{Model development and iteration}
\label{sec:model_development}

We developed the pipeline on diarrhea, our \emph{development set}, because
expert annotation of all 18~phenotypes across every patient would be
prohibitively slow, the bottleneck this pipeline relieves.

The architecture was settled first, by a comparison of six extraction
strategies on this development set that selected the dual-model consensus gate.
Holding it fixed, we then ran that consensus configuration under five
classification prompts differing only in what they count as diarrhea, alongside
two rule-based methods that bound the comparison from below
(Table~\ref{tab:model_comparison}). Prompt wording, the architecture
comparison, and the identifiers these methods carry in the released code are in
Supplementary Notes~S2, S3, and~S7.

We selected the reported prompt on the same 183 cells we score it on. Its diarrhea
accuracy is therefore an in-sample upper bound rather than a held-out estimate. The
evidence that the design generalizes comes from the other 17 phenotypes. We froze
the pipeline after the diarrhea work and applied it to them with only the phenotype
definition substituted (Results; Supplementary Note~S2).

\subsection*{Ethical considerations}

This study was conducted under IRB protocol H-52222 at Baylor College of
Medicine / Texas Children's Hospital. All data remained within the
institutional secure computing environment, and LLM inference was performed
in-environment under enterprise data-governance controls.

\section*{Results}
\label{sec:results}

\subsection*{Contextual reasoning improves performance and outperforms keyword matching}
\label{sec:diarrhea_validation}

We evaluated seven methods of increasing specificity on the identical
183-cell ground truth (Table~\ref{tab:model_comparison}).

\paragraph{The severe-diarrhea prompt performs best.}
Accuracy rises with the severity threshold the prompt sets, and the
severe-diarrhea prompt (Claude~Opus~4.6) performs best on every headline
metric except recall: 84.2\% accuracy, precision 0.714, F1 0.674, $\kappa$ 0.570.
This is 47 percentage points over the regular-expression baseline (37.2\%,
indiscriminate over-flagging) and 42 points over negation-aware NLP (42.1\%). Both
restrictions help. Requiring diarrhea to be uncontrolled improves on counting any
documented mention (79.6\% against 75.2\%), and stating that restriction as an
explicit severity bar raises accuracy to 84.2\% via precision
(0.596$\rightarrow$0.714), as the bar converts nine false positives into true
negatives at the cost of one true positive.

\paragraph{How strong is this evidence?}
We then asked whether these differences are real or only directional, testing each
comparison with McNemar's exact test and Holm-correcting across a pre-declared
family of seven prompt comparisons (Supplementary Table~S4). The reported prompt
beats both rule-based methods decisively ($p<0.0001$), and every prompt sits far
above them on accuracy. None of the prompt-to-prompt comparisons reaches
significance. The severe-diarrhea prompt wins 9 cells against
the uncontrolled-diarrhea prompt the pipeline was originally deployed with (row 5 of
Table~\ref{tab:model_comparison}) and loses 1, which is $p=0.021$ before correction and $p=0.15$
after, so it is the top of a sweep rather than a demonstrated improvement. With 183
cells and a gap of roughly 8, this cohort is too small to resolve differences of
that size.

\begin{table}[H]
\centering
\caption{Method comparison on the diarrhea phenotype. All seven methods are
scored on the identical 183-cell expert ground truth from the 56 labeled patients. \emph{Type} is
the kind of system doing the reading, either lexical pattern matching, rule-based
NLP, or a large language model. The five LLM rows share the same two agents, the
same consensus gate, and the same model (Claude~Opus~4.6), and differ only in what
they count as diarrhea. All five also exclude resolved, negated, templated,
anticipatory and other-cause mentions, so the two \emph{with rules} variants restate
that guidance rather than introduce it. Accuracy is the mean$\pm$SD over three independent
runs, and the two non-LLM methods are deterministic. The other metrics are
computed from the modal prediction across the three runs. The bolded row is the configuration reported throughout the paper.
Full prompt wording, and the identifiers these names carry in the released code,
are in Supplementary Notes~S3 and~S7. Best value in each column in bold.}
\label{tab:model_comparison}
\setlength{\tabcolsep}{5pt}
\begin{tabular}{p{3.0cm}cp{4.3cm}cccc}
\toprule
\textbf{Method} & \textbf{Type} & \textbf{What counts as diarrhea} & \textbf{Accuracy (\%)} & \textbf{Precision} & \textbf{Recall} & \textbf{F1} \\
\midrule
Regular expression & Regex &
Any note containing a diarrhea keyword &
37.2 & 0.282 & \textbf{0.936} & 0.434 \\
\addlinespace
Negation-aware NLP & NLP &
The same keywords, with negated and historical mentions filtered out &
42.1 & 0.288 & 0.851 & 0.430 \\
\addlinespace
Any diarrhea, with rules & LLM &
Any documented diarrhea, unless the note reports it as resolved or attributes it
to another cause &
75.0$\pm$0.3 & 0.507 & 0.766 & 0.610 \\
\addlinespace
Any diarrhea & LLM &
Any documented diarrhea, including diarrhea that treatment has brought under
control &
75.2$\pm$0.3 & 0.514 & 0.787 & 0.622 \\
\addlinespace
Uncontrolled diarrhea & LLM &
Diarrhea unless the note describes it as mild or controlled without needing
intervention &
79.6$\pm$0.3 & 0.596 & 0.660 & 0.626 \\
\addlinespace
Severe diarrhea, with rules & LLM &
Active diarrhea causing concern or prompting a management change, unless the note
reports it as resolved or attributes it to another cause &
83.1$\pm$0.5 & 0.700 & 0.596 & 0.644 \\
\addlinespace
\textbf{Severe diarrhea} & \textbf{LLM} &
\textbf{Diarrhea that is active and either causing ongoing concern or prompting
a change in management at that visit} &
\textbf{84.2$\pm$0.0} & \textbf{0.714} & 0.638 & \textbf{0.674} \\
\bottomrule
\end{tabular}
\end{table}

A patient-level bootstrap (2{,}000 resamples) gave a 95\% CI of 78.1--89.9\%, and
splitting the cohort in half by patient gives 81.1\% and 87.5\%. Restricted to the
129 on-treatment cells accuracy is 79.1\%, because the 54 baseline cells are almost
all true negatives and inflate every configuration equally (baseline 96.3\%), with
the ranking unchanged (Supplementary Note~S7).

The predicted vs.\ ground-truth prevalence trajectories
(Figure~\ref{fig:diarrhea}C) show the model \emph{under}-detecting at 6~months
(45.5\% vs.\ 63.6\%) and \emph{over}-detecting at 24~months (30.8\% vs.\
15.4\%), with the remaining timepoints within a few points of the expert labels,
attributed to a stricter human threshold for ``managed'' diarrhea and to
documentation falling one window away from the labeled visit, dominated by
annotation and data-availability factors rather than model misreadings.
Per-timepoint confusion matrices and additional metrics are in Supplementary
Tables~S5 and~S6.

\subsection*{The pipeline extends to 18 phenotypes over 30 months of follow-up}
\label{sec:multi_phenotype}

We extended the analysis to the other 17 phenotypes using a uniform presence
prompt, with documented presence shown in Figure~\ref{fig:multipheno} and full
per-visit rates in Supplementary Tables~S7--S8.
Because diarrhea has its own ground truth and error analysis, we assessed these
17 with a lighter expert spot-check instead. For each
we drew a balanced sample of six cases, three model-\emph{present} and three
model-\emph{absent}, and the Rett Center clinical
team judged whether the model's call agreed with the note. The model agreed with
the reviewer in \textbf{90.2\% of cases} (92/102; 95\% CI 82.9--94.6\%), 86.3\%
on model-positive calls (44/51) and 94.1\% on model-negative (48/51), the
pipeline slightly more conservative about missing than over-calling a
phenotype. The review procedure and the scope of the reviewed sample are in
Supplementary Note~S6.

Per-phenotype agreement (Table~\ref{tab:trust_ratings}) shows nine phenotypes
reached 6/6, six reached 5/6, and two (dysautonomia, refusing medication) reached
4/6. One pattern dominated the ten disagreements, in which the model inferred a
currently active phenotype from a historical or past-medical-history
mention. Attention/alertness
and seizure, whose vocabulary overlaps routine documentation and required careful
presence-specific keyword design, both nonetheless reached 6/6. Inter-LLM
agreement per phenotype provides a complementary label-free check
(Supplementary Note~S1, Supplementary Figure~S1).

\begin{table}[H]
\centering
\caption{Expert spot-check agreement for the 17 phenotypes without a ground
truth. For each phenotype the Rett Center clinical team judged six
model calls (three model-positive and three model-negative) as agreeing or
disagreeing with the note (102 cases in total); overall model--reviewer
agreement was 90.2\% (92/102; 95\% CI 82.9--94.6\%). Diarrhea's six cases
were omitted from this spot-check because diarrhea is validated against a
quantitative ground truth and characterized by a dedicated error analysis.}
\label{tab:trust_ratings}
\begin{tabular}{clp{7.4cm}}
\toprule
\textbf{Agreement} & \textbf{Count} & \textbf{Phenotypes} \\
\midrule
6/6 (100\%) & 9 & Aggressive Behavior, Attention/Alertness,
Breathing/Hyperventilation, Fine Motor, Gross Motor, Language/Communication,
Seizure, Stereotypy, Vomiting \\
\addlinespace
5/6 (83\%) & 6 & Anxiety, Bruxism, General Mood, Other Side Effects, Sleep,
Weight Loss \\
\addlinespace
4/6 (67\%) & 2 & Dysautonomia, Refusing Medication \\
\bottomrule
\end{tabular}
\end{table}

\subsection*{Adverse-effect and clinical-feature presence follow distinct temporal patterns}
\label{sec:clinical_findings}

Several patterns emerge across the extracted phenotypes
(Figure~\ref{fig:multipheno}).

\paragraph{Adverse-effect presence.}
Documented rates vary across the adverse effects (Figure~\ref{fig:multipheno}A).
Other side effects are documented in 11\% of patients at baseline and rise
sharply to 58\% at 6~months, settling at 38--49\% thereafter. Medication refusal
rests on a single patient at baseline, so that cell is suppressed, and it peaks
at 38\% at 18~months before falling to 15--16\% at the later visits.
Weight-loss documentation is already high before treatment starts (63\%) and
stays in a 62--71\% band across follow-up, so it is unlikely to be
treatment-emergent and more consistent with a pre-existing feeding or growth
concern. Vomiting fluctuates modestly, peaking at 38\% at 12~months and falling
to 22\% by 30~months. Diarrhea,
the only validated adverse effect, is characterized separately
(Figure~\ref{fig:diarrhea}).

\paragraph{Rett clinical-feature presence.}
These 13 features are manifestations of Rett syndrome itself and not adverse
effects of trofinetide, so their documented presence describes the disorder
under treatment rather than harm caused by it. They span a wide range of
documented prevalence
(Figure~\ref{fig:multipheno}A; full rates in Supplementary Table~S8). Gross- and
fine-motor impairment, impaired language/communication, and hand stereotypies are
the core neurodevelopmental features of the disorder, and all are documented as
present in nearly all patients at every visit (92--100\%).
Dysautonomia, breathing, bruxism, sleep, and seizure form a middle band at
intermediate rates, while attention/alertness, anxiety, and the affective and
behavioral findings of mood and aggression are documented least often.
Documented presence at the pre-treatment baseline is lower than at 6 months for
5 of the 13 features, so baseline-to-follow-up differences should still not be
read as clinical change (Supplementary Table~S8). These
trajectories are documentation-based and, except for diarrhea, are not
validated against a ground truth (Discussion).

\subsection*{Most residual discrepancies reflect differing scoring targets rather than misreading of the note}
\label{sec:error_analysis}

We classified every diarrhea discrepancy under the reported configuration (29 of
183 cells) by root cause (Table~\ref{tab:error_types},
Figure~\ref{fig:diarrhea}D), assigning each cell by rule from the
keyword-passing notes in its own window and the adjacent ones rather than by
manual re-reading. Three mechanisms account for all 29, and all three are
annotation, data-coverage, or window-placement limitations rather than
demonstrable misreadings. The reviewer integrates patient-level context
across encounters and grades case by case, while the model is note-local by
construction and applies one fixed threshold uniformly.

\begin{table}[H]
\centering
\caption{Error taxonomy for the diarrhea phenotype under the reported
configuration (29 total errors on 183 cells, 17 false negatives and 12 false
positives). ``Model reading defensible?'' indicates whether the model's call is
defensible against the notes available in the window it scored. Assignment is
rule-based, so it separates the window a piece of evidence falls in but not
whether a note was misread within its own window.}
\label{tab:error_types}
\begin{tabular}{p{4.3cm}p{6.0cm}cc}
\toprule
\textbf{Mechanism} & \textbf{What happens} & \textbf{N (\%)} & \textbf{Model reading defensible?} \\
\midrule
Threshold disagreement &
Note documents active or medication-managed diarrhea. The reviewer scores No under a
stricter severity threshold, or the converse, where the reviewer
counts mild or managed changes the model discounts &
14 (48\%) & Arguable \\
\addlinespace
Temporal displacement &
The supporting positive note exists but falls in an adjacent six-month window, so
the model and the reviewer are describing the same episode at different visits &
12 (41\%) & Arguable \\
\addlinespace
Evidence absent &
Diarrhea recorded in the ground truth but not present in any keyword-passing note
in the window, having been reported by phone, through the patient portal, or only
retrospectively &
3 (10\%) & N/A (no evidence) \\
\bottomrule
\end{tabular}
\end{table}

Most discrepancies are not misreadings. The largest category (14/29, 48\%) is
a \emph{definitional} disagreement about what counts as clinically
significant diarrhea rather than a model error. It cuts both ways, since the 29
errors are 17 false negatives and 12 false positives, so a single severity
threshold would trade one error type for the other.
Re-scoring the 14 threshold-disagreement cases as correct yields a 91.8\%
bounding sensitivity-analysis estimate rather than a corrected performance
figure (Supplementary Note~S4).

The 12 displacement cells are the ones an onset-aligned variant of the
predictions addresses (Supplementary Note~S7). Errors
concentrate at 6~months, which holds 18 of the 29, and 28 of the 29 were also
errors under the deployed prompt, so the severity bar shrank the
error set rather than relocating it. What this rule-based assignment cannot do is
separate a within-note misreading from a window-placement mismatch, since the
prompts emit only YES or NO and so record whether a call is right but not the
reasoning behind it (Supplementary Note~S4).

\section*{Discussion}
\label{sec:discussion}

A dual-LLM consensus pipeline can extract clinically meaningful
treatment-response phenotypes from unstructured notes at scale, recovering signal
that structured fields miss. Modeling clinical context significantly outperforms
naive keyword matching, and what separates the configurations that do so is what
the prompt counts as diarrhea rather than how many models are asked. The
severe-diarrhea prompt was strongest, but its margin over the deployed prompt is
unresolved at this sample size, so we adopt it as a precision-favorable default
rather than a proven improvement. It was also selected on the same 183 cells it is
scored on, so its accuracy is an in-sample upper bound. The evidence that the
design transfers is the half-cohort split and the 90.2\% agreement on the other 17
phenotypes, which the same architecture and prompt design reached with only the
phenotype definition substituted.

The type of error matters more than the headline accuracy. Most diarrhea errors
are annotation or data limitations rather than model failures, which tells a
deploying institution which discrepancies prompting can fix, which are
data-coverage ceilings, and which are annotation-guideline decisions. Design
choices and intended use are in Supplementary Note~S4, and the architecture
comparison is in Note~S2.

Diarrhea is the most commonly reported adverse event in the pivotal Phase~3
trial, at ${\sim}$80\% in the treatment
arm~\cite{neul2023trofinetide,abbas2024safety}. Our validated target is
narrower, severe rather than any diarrhea, and expert labels put it at 64\% of
evaluable patients at 6~months and below 20\% at every later visit
(Supplementary Table~S5), so the difference from the trial figure is mostly one
of threshold rather than of detection. The later validated timepoints have few evaluable cells, 13 at 24~months
and 6 at 30~months, and warrant caution.

The main limitation is that only diarrhea has a ground truth. The other 17
phenotypes rest on spot-check review and inter-LLM agreement rather than full
annotation (Supplementary Note~S1), a resource constraint described in Methods.
Their rates are therefore hypothesis-generating lower bounds, and generalization
likely varies by phenotype, since concrete adverse effects should transfer more
readily than subjective findings. The pipeline detects only what our corpus documents, and some
symptoms surface only in other channels. The exposure anchor is adjudicated from
note text for most patients rather than taken from a structured record, so window
boundaries carry real uncertainty, visible in the 12 displacement errors. The observed trajectories cannot be definitively
attributed to trofinetide rather than to concurrent therapies or the natural
course of development. The outcome is documented presence and not direction of
change or severity, so the pipeline cannot distinguish an improved feature from a
persistent one. For the core Rett features, documented in nearly all patients at
nearly every visit, detecting within-feature improvement would need
severity-graded instruments such as the RSBQ or CGI-I. Documentation does move for
other side effects and refusing medication, so the method registers change when
change is recorded. All patients come from a single institution, and the pipeline
is not yet optimized for efficiency, which cheap pre-filtering, batched calls, and
smaller distilled models could improve.

Integrating additional sources such as telephone logs and portal messages, and
cross-temporal reasoning that propagates retrospective mentions to the timepoint
they describe, would recover symptoms currently missed. Expanding the ground truth
by second-reviewer adjudication and by annotation prioritized on inter-LLM
disagreement would convert the six-case agreement tiers into per-phenotype
accuracy with confidence intervals. Individual feature trajectories, including the
apparent seizure rise, need validated per-timepoint annotation before clinical
interpretation and are deferred. Direction-of-change extraction is feasible in the
same architecture with phenotype-specific prompts and a correspondingly annotated
ground truth, but was outside this validation's scope. The framework extends to
other medications and conditions by configuration alone.

\section*{Conclusion}
\label{sec:conclusion}

We present a dual-LLM consensus pipeline for multi-phenotype treatment-response
extraction from clinical notes, applied to 108 patients with Rett syndrome receiving
trofinetide over 30 months. Using Claude~Opus~4.6, it substantially outperforms
a regular-expression baseline on an expert-annotated diarrhea development set, and the
same configuration transfers out-of-sample to the remaining 17 phenotypes with only
the phenotype definition substituted.
It also yields one of the first comprehensive real-world characterizations of
how multi-domain trofinetide response is documented in routine care. The architecture extends to further
phenotypes, medications, and institutions by configuration alone, showing that
LLM consensus can bridge clinical narrative and the structured data needed for
scalable treatment-response monitoring and real-world evidence in rare
diseases.



\section*{Funding}
This research was supported by a fellowship from the Gulf Coast Consortia on
the NLM Training Program in Biomedical Informatics and Data Science
(T15~LM007093); the National Science Foundation Graduate Research
Fellowship Program (NSF GRFP Fellow ID 2024370642); the Cancer Prevention and Research
Institute of Texas (CPRIT, RP240131); the Chan Zuckerberg Initiative
(2023-332162); the National Institutes of Health (NIH, U54NS093793 and
OT2OD040565); the Eunice Kennedy Shriver National Institute of Child Health
and Human Development of the NIH (P50HD103555); the Chao Endowment; the
Huffington Foundation; and the Jan and Dan Duncan Neurological Research
Institute at Texas Children's Hospital.
D.Pe.\ is supported by Doris Duke Charitable Foundation grant 2023-0235 and NINDS K23 NS125126-01A1.

\section*{Additional Contributions}
No AI-assisted tools were used for study design, analysis, or primary
manuscript drafting; generative AI tools were used only for proofreading and
typographical/grammatical correction of author-written text. Large language
models are the object of study and their use in the analysis pipeline is
described in Methods. In accordance with COPE's position and JAMIA policy, no AI
or NLP tool is listed as an author; the authors reviewed and verified all text
and take full responsibility for the integrity, accuracy, and originality of all
content.

\section*{Conflicts of Interest}
D.Pe.\ provides paid consulting services for Ionis Pharmaceuticals, M2DS Therapeutics and Acadia Pharmaceuticals. B.S.\ has provided consulting services to Acadia, Neurogene, Ionis Pharmaceuticals, Taysha, EcoR1, and UCB; all remuneration for these has been made to his institution. The remaining authors declare no competing interests.

\section*{Data Availability}
The classification pipeline, phenotype keyword lists, and LLM prompt templates
are publicly available at \url{https://github.com/DanPalBCM/clinical-trofinetide}.
The underlying clinical notes cannot be shared due to
patient-privacy protections and institutional/HIPAA policy; requests for access
to de-identified derived data may be directed to the corresponding author and
are subject to a data-use agreement and IRB approval. De-identified aggregate
results supporting all figures and tables are provided within the article and
its supplementary materials. Figures were produced with Matplotlib in a
soft-fill style inspired by the PubliPlots library~\cite{botas2025publiplots}.
Reporting follows the TRIPOD-LLM guidance for studies using large language
models on clinical text~\cite{gallifant2025tripod};
the addressed checklist items are mapped to their locations in Supplementary
Note~S5.

\section*{Author Contributions}
Following the CRediT taxonomy:
\textbf{D.P. (Palacios):} conceptualization, methodology, software,
formal analysis, investigation, data curation, visualization,
writing -- original draft.
\textbf{G.Q. (Qi):} visualization (Figure~1A),
writing -- original draft.
\textbf{M.G.K. (Guzman~Karlsson):} conceptualization, investigation,
methodology (bioinformatics expertise, analysis strategy),
writing -- review and editing.
\textbf{S.K.D. (David):} investigation, data curation, validation (registry
curation, diarrhea ground-truth annotation, and multi-phenotype spot-check).
\textbf{M.S. (Sameeruddin):} investigation, data curation, validation (registry
curation, diarrhea ground-truth annotation, and multi-phenotype spot-check).
\textbf{B.S. (Suter):} investigation, data curation, validation (registry
curation, diarrhea ground-truth annotation, and multi-phenotype spot-check),
clinical/domain expertise, writing -- review and editing.
\textbf{Z.L. (Liu):} supervision, funding acquisition, methodology,
writing -- review and editing.
\textbf{D.Pe. (Pehlivan):} investigation, data curation, validation (registry
curation, diarrhea ground-truth annotation, and multi-phenotype spot-check),
clinical/domain expertise, resources (patient registry),
writing -- review and editing.
\textbf{H.C. (Chen):} supervision, project administration, methodology,
writing -- review and editing.
All authors reviewed and approved the final manuscript.

\bibliography{references}

@article{amir1999rett,
  title={Rett syndrome is caused by mutations in X-linked MECP2, encoding methyl-CpG-binding protein 2},
  author={Amir, Ruthie E and Van den Veyver, Ignatia B and Wan, Mimi and Tran, Charles Q and Francke, Uta and Zoghbi, Huda Y},
  journal={Nature Genetics},
  volume={23},
  number={2},
  pages={185--188},
  year={1999},
  doi={10.1038/13810},
  publisher={Nature Publishing Group}
}

@article{neul2010rett,
  title={Rett syndrome: revised diagnostic criteria and nomenclature},
  author={Neul, Jeffrey L and Kaufmann, Walter E and Glaze, Daniel G and Christodoulou, John and Clarke, Angus J and Bahi-Buisson, Nadia and Leonard, Helen and Bailey, Mark E S and Schanen, N Carolyn and Zappella, Michele and Renieri, Alessandra and Huppke, Peter and Percy, Alan K},
  journal={Annals of Neurology},
  volume={68},
  number={6},
  pages={944--950},
  year={2010},
  doi={10.1002/ana.22124},
  publisher={Wiley}
}

@article{glaze2017double,
  title={A double-blind, randomized, placebo-controlled clinical study of trofinetide in the treatment of Rett syndrome},
  author={Glaze, Daniel G and Neul, Jeffrey L and Percy, Alan and Feyma, Tim and Beisang, Arthur and Yaroshinsky, Alex and Stoms, George and Zuchero, David and Horrigan, Joseph and Glass, Larry and Jones, Nancy E},
  journal={Pediatric Neurology},
  volume={76},
  pages={37--46},
  year={2017},
  doi={10.1016/j.pediatrneurol.2017.07.002},
  note={Phase 2, RETT-001; adolescent and adult females},
  publisher={Elsevier}
}

@article{neul2023trofinetide,
  title={Trofinetide for the treatment of Rett syndrome: a randomized phase 3 study},
  author={Neul, Jeffrey L and Percy, Alan K and Benke, Timothy A and Berry-Kravis, Elizabeth M and Glaze, Daniel G and Marsh, Eric D and Lin, Tim and Stankovic, Serge and Bishop, Kathie M and Youakim, James M},
  journal={Nature Medicine},
  volume={29},
  number={6},
  pages={1468--1475},
  year={2023},
  doi={10.1038/s41591-023-02398-1},
  note={LAVENDER, NCT04181723},
  publisher={Nature Publishing Group}
}

@misc{fda2023daybue,
  title   = {{FDA} Approves First Treatment for {R}ett Syndrome},
  author  = {{U.S. Food and Drug Administration}},
  year    = {2023},
  url     = {https://www.fda.gov/drugs/news-events-human-drugs/fda-approves-first-treatment-rett-syndrome},
  note    = {Content current as of 13 March 2023. Accessed 2026-08-18}
}

@article{abbas2024safety,
  title={Safety and efficacy of trofinetide in Rett syndrome: a systematic review and meta-analysis of randomized controlled trials},
  author={Abbas, Abdallah and Fayoud, Aya M and El Din Moawad, Mostafa Hossam and Hamad, Abdullah Ashraf and Hamouda, Heba and Fouad, Eman A},
  journal={BMC Pediatrics},
  volume={24},
  number={1},
  pages={206},
  year={2024},
  doi={10.1186/s12887-024-04526-3},
  publisher={Springer}
}

@article{savova2010mayo,
  title={Mayo clinical Text Analysis and Knowledge Extraction System ({cTAKES}): architecture, component evaluation and applications},
  author={Savova, Guergana K and Masanz, James J and Ogren, Philip V and Zheng, Jiaping and Sohn, Sunghwan and Kipper-Schuler, Karin C and Chute, Christopher G},
  journal={Journal of the American Medical Informatics Association},
  volume={17},
  number={5},
  pages={507--513},
  year={2010},
  doi={10.1136/jamia.2009.001560},
  publisher={Oxford University Press}
}

@article{fu2020clinical,
  title={Clinical concept extraction: a methodology review},
  author={Fu, Sunyang and Chen, David and He, Huan and Liu, Sijia and Moon, Sungrim and Peterson, Kevin J and Shen, Feichen and Wang, Liwei and Wang, Yanshan and Wen, Andrew and Zhao, Yiqing and Sohn, Sunghwan and Liu, Hongfang},
  journal={Journal of Biomedical Informatics},
  volume={109},
  pages={103526},
  year={2020},
  doi={10.1016/j.jbi.2020.103526},
  publisher={Elsevier}
}

@article{shivade2014review,
  title={A review of approaches to identifying patient phenotype cohorts using electronic health records},
  author={Shivade, Chaitanya and Raghavan, Preethi and Fosler-Lussier, Eric and Embi, Peter J and Elhadad, Noemie and Johnson, Stephen B and Lai, Albert M},
  journal={Journal of the American Medical Informatics Association},
  volume={21},
  number={2},
  pages={221--230},
  year={2014},
  doi={10.1136/amiajnl-2013-001935},
  publisher={Oxford University Press}
}

@article{kirby2016phekb,
  title={{PheKB}: a catalog and workflow for creating electronic phenotype algorithms for transportability},
  author={Kirby, Jacqueline C and Speltz, Peter and Rasmussen, Luke V and Basford, Melissa and Gottesman, Omri and Peissig, Peggy L and Pacheco, Jennifer A and Tromp, Gerard and Pathak, Jyotishman and Carrell, David S and Ellis, Stephen B and Lingren, Todd and Thompson, Will K and Savova, Guergana and Haines, Jonathan and Roden, Dan M and Harris, Paul A and Denny, Joshua C},
  journal={Journal of the American Medical Informatics Association},
  volume={23},
  number={6},
  pages={1046--1052},
  year={2016},
  doi={10.1093/jamia/ocv202},
  publisher={Oxford University Press}
}

@article{wang2018clinical,
  title={Clinical information extraction applications: a literature review},
  author={Wang, Yanshan and Wang, Liwei and Rastegar-Mojarad, Majid and Moon, Sungrim and Shen, Feichen and Afzal, Naveed and Liu, Sijia and Zeng, Yuqun and Mehrabi, Saeed and Sohn, Sunghwan and Liu, Hongfang},
  journal={Journal of Biomedical Informatics},
  volume={77},
  pages={34--49},
  year={2018},
  doi={10.1016/j.jbi.2017.11.011},
  publisher={Elsevier}
}

@article{kreimeyer2017natural,
  title={Natural language processing systems for capturing and standardizing unstructured clinical information: a systematic review},
  author={Kreimeyer, Kory and Foster, Matthew and Pandey, Abhishek and Arya, Nina and Halford, Gwendolyn and Jones, Sandra F and Forshee, Richard and Walderhaug, Mark and Botsis, Taxiarchis},
  journal={Journal of Biomedical Informatics},
  volume={73},
  pages={14--29},
  year={2017},
  doi={10.1016/j.jbi.2017.07.012},
  publisher={Elsevier}
}

@article{harkema2009context,
  title={{ConText}: an algorithm for determining negation, experiencer, and temporal status from clinical reports},
  author={Harkema, Henk and Dowling, John N and Thornblade, Tyler and Chapman, Wendy W},
  journal={Journal of Biomedical Informatics},
  volume={42},
  number={5},
  pages={839--851},
  year={2009},
  doi={10.1016/j.jbi.2009.05.002},
  publisher={Elsevier}
}

@article{chapman2001negex,
  title={A simple algorithm for identifying negated findings and diseases in discharge summaries},
  author={Chapman, Wendy W and Bridewell, Will and Hanbury, Paul and Cooper, Gregory F and Buchanan, Bruce G},
  journal={Journal of Biomedical Informatics},
  volume={34},
  number={5},
  pages={301--310},
  year={2001},
  doi={10.1006/jbin.2001.1029},
  publisher={Elsevier}
}

@article{murff2011automated,
  title={Automated identification of postoperative complications within an electronic medical record using natural language processing},
  author={Murff, Harvey J and FitzHenry, Fern and Matheny, Michael E and Gentry, Nancy and Kotter, Kristen L and Crimin, Kimberly and Dittus, Robert S and Rosen, Amy K and Elkin, Peter L and Brown, Steven H and Speroff, Theodore},
  journal={JAMA},
  volume={306},
  number={8},
  pages={848--855},
  year={2011},
  doi={10.1001/jama.2011.1204},
  publisher={American Medical Association}
}

@article{harpaz2012novel,
  title={Novel data-mining methodologies for adverse drug event discovery and analysis},
  author={Harpaz, Rave and DuMouchel, William and Shah, Nigam H and Madigan, David and Ryan, Patrick and Friedman, Carol},
  journal={Clinical Pharmacology \& Therapeutics},
  volume={91},
  number={6},
  pages={1010--1021},
  year={2012},
  doi={10.1038/clpt.2012.50},
  publisher={Wiley}
}

@article{singhal2023large,
  title={Large language models encode clinical knowledge},
  author={Singhal, Karan and Azizi, Shekoofeh and Tu, Tao and Mahdavi, S Sara and Wei, Jason and Chung, Hyung Won and Scales, Nathan and Tanwani, Ajay and Cole-Lewis, Heather and Pfohl, Stephen and others},
  journal={Nature},
  volume={620},
  number={7972},
  pages={172--180},
  year={2023},
  doi={10.1038/s41586-023-06291-2},
  publisher={Nature Publishing Group}
}

@article{thirunavukarasu2023large,
  title={Large language models in medicine},
  author={Thirunavukarasu, Arun James and Ting, Darren Shu Jeng and Elangovan, Kabilan and Gutierrez, Laura and Tan, Ting Fang and Ting, Daniel Shu Wei},
  journal={Nature Medicine},
  volume={29},
  number={8},
  pages={1930--1940},
  year={2023},
  doi={10.1038/s41591-023-02448-8},
  publisher={Nature Publishing Group}
}

@misc{nori2023capabilities,
  title={Capabilities of {GPT-4} on Medical Challenge Problems},
  author={Nori, Harsha and King, Nicholas and McKinney, Scott Mayer and Carignan, Dean and Horvitz, Eric},
  year={2023},
  eprint={2303.13375},
  archivePrefix={arXiv},
  primaryClass={cs.CL},
  doi={10.48550/arXiv.2303.13375},
  url={https://arxiv.org/abs/2303.13375}
}

@misc{liu2023deid,
  title={{DeID-GPT}: Zero-shot Medical Text De-Identification by {GPT-4}},
  author={Liu, Zhengliang and Yu, Xiaowei and Zhang, Lu and Wu, Zihao and Cao, Chao and Dai, Haixing and Zhao, Lin and Liu, Wei and Shen, Dinggang and Li, Quanzheng and Liu, Tianming and Zhu, Dajiang and Li, Xiang},
  year={2023},
  eprint={2303.11032},
  archivePrefix={arXiv},
  primaryClass={cs.CL},
  doi={10.48550/arXiv.2303.11032},
  url={https://arxiv.org/abs/2303.11032},
  note={arXiv:2303.11032v1}
}

@misc{tang2023does,
  title={Does Synthetic Data Generation of {LLMs} Help Clinical Text Mining?},
  author={Tang, Ruixiang and Han, Xiaotian and Jiang, Xiaoqian and Hu, Xia},
  year={2023},
  eprint={2303.04360},
  archivePrefix={arXiv},
  primaryClass={cs.CL},
  doi={10.48550/arXiv.2303.04360},
  url={https://arxiv.org/abs/2303.04360}
}

@misc{gu2025surveyllmasajudge,
  title={A Survey on {LLM}-as-a-Judge},
  author={Gu, Jiawei and Jiang, Xuhui and Shi, Zhichao and Tan, Hexiang and Zhai, Xuehao and Xu, Chengjin and Li, Wei and Shen, Yinghan and Ma, Shengjie and Liu, Honghao and Wang, Saizhuo and Zhang, Kun and Wang, Yuanzhuo and Gao, Wen and Ni, Lionel and Guo, Jian},
  year={2024},
  eprint={2411.15594},
  archivePrefix={arXiv},
  primaryClass={cs.CL},
  doi={10.48550/arXiv.2411.15594},
  url={https://arxiv.org/abs/2411.15594}
}

@misc{mcaleese2024critics,
  title={{LLM} Critics Help Catch {LLM} Bugs},
  author={McAleese, Nat and Pokorny, Rai Michael and Ceron Uribe, Juan Felipe and Nitishinskaya, Evgenia and Trebacz, Maja and Leike, Jan},
  year={2024},
  eprint={2407.00215},
  archivePrefix={arXiv},
  primaryClass={cs.SE},
  doi={10.48550/arXiv.2407.00215},
  url={https://arxiv.org/abs/2407.00215}
}

@article{gallifant2025tripod,
  title={The {TRIPOD-LLM} reporting guideline for studies using large language models},
  author={Gallifant, Jack and Afshar, Majid and Ameen, Saleem and Aphinyanaphongs, Yindalon and Chen, Shan and Cacciamani, Giovanni and Demner-Fushman, Dina and Dligach, Dmitriy and Daneshjou, Roxana and Fernandes, Chrystinne and others},
  journal={Nature Medicine},
  volume={31},
  number={1},
  pages={60--69},
  year={2025},
  doi={10.1038/s41591-024-03425-5},
  publisher={Nature Publishing Group}
}

@misc{wang2024rethinkingboundsllmreasoning,
  title         = {Rethinking the Bounds of {LLM} Reasoning: Are Multi-Agent
                   Discussions the Key?},
  author        = {Wang, Qineng and Wang, Zihao and Su, Ying and
                   Tong, Hanghang and Song, Yangqiu},
  year          = {2024},
  eprint        = {2402.18272},
  archivePrefix = {arXiv},
  primaryClass  = {cs.CL},
  url           = {https://arxiv.org/abs/2402.18272}
}

@misc{cemri2025multiagentllmsystemsfail,
  title         = {Why Do Multi-Agent {LLM} Systems Fail?},
  author        = {Cemri, Mert and Pan, Melissa Z. and Yang, Shuyi and
                   Agrawal, Lakshya A. and Chopra, Bhavya and
                   Tiwari, Rishabh and Keutzer, Kurt and
                   Parameswaran, Aditya and Klein, Dan and
                   Ramchandran, Kannan and Zaharia, Matei and
                   Gonzalez, Joseph E. and Stoica, Ion},
  year          = {2025},
  eprint        = {2503.13657},
  archivePrefix = {arXiv},
  primaryClass  = {cs.AI},
  url           = {https://arxiv.org/abs/2503.13657}
}

@misc{botas2025publiplots,
  title        = {{PubliPlots}: Publication-ready plotting for {Python}},
  author       = {Botas, Jorge},
  year         = {2025},
  howpublished = {\url{https://github.com/jorgebotas/publiplots}}
}

\clearpage
\section*{Figures}

\begin{figure}[htbp]
\centering
\includegraphics[width=\textwidth]{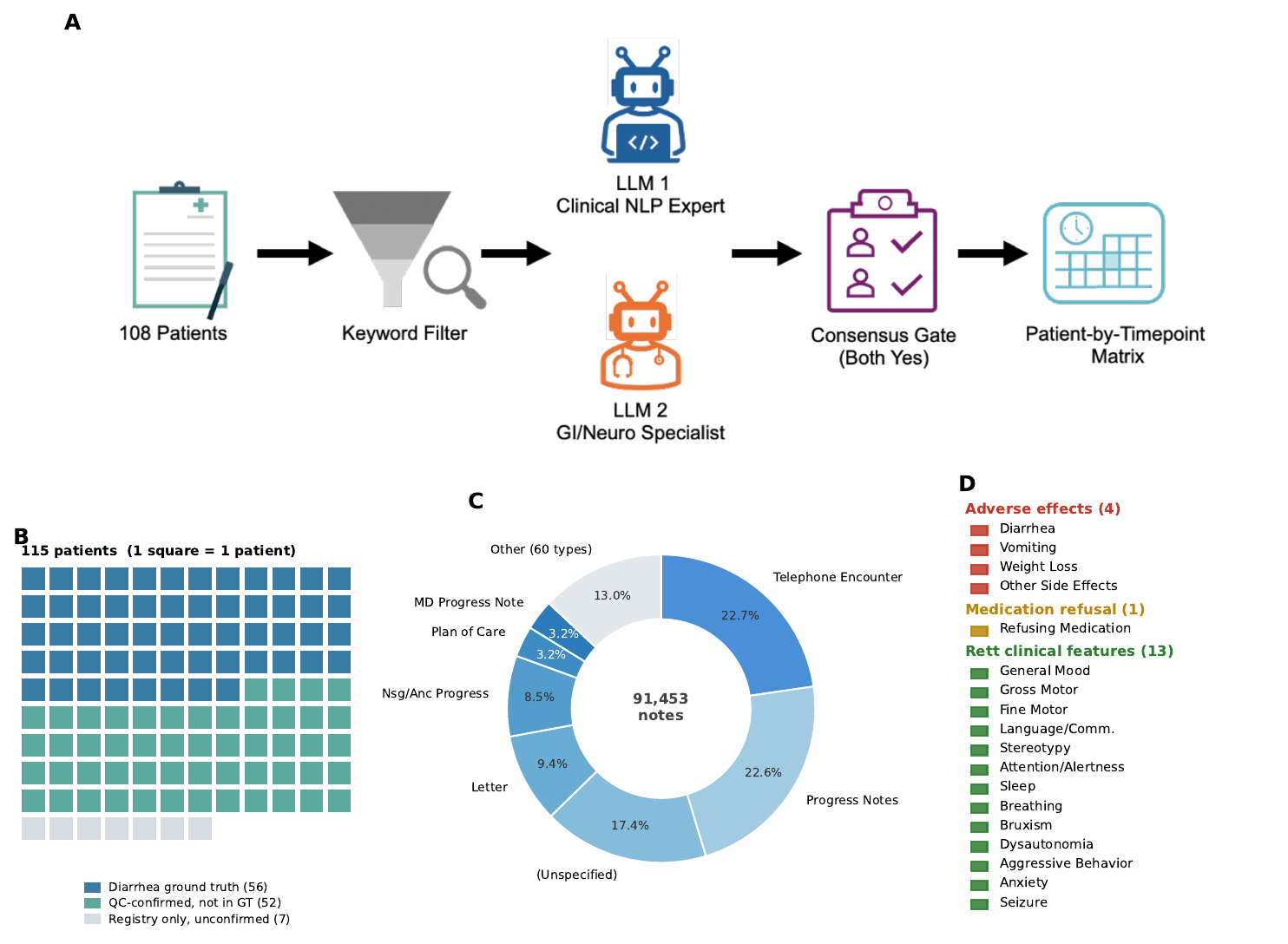}
\caption{Study design, cohort, corpus, and phenotype panel.
\textbf{(A)}~The five-stage extraction pipeline, from keyword filtering through
independent classification by two LLMs, the consensus gate, aggregation into a
patient-by-timepoint matrix, and evaluation against ground truth.
\textbf{(B)}~The cohort as an icon array, one square per patient, showing the 115
patients screened, the 108 with QC-confirmed trofinetide exposure, and the 56 of
those who formed the expert-annotated diarrhea ground truth.
\textbf{(C)}~Composition of the 91,453 retrieved notes by note type, with the top
seven shown and 60 further types collapsed into ``Other''.
\textbf{(D)}~The 18 extracted phenotypes, grouped into adverse effects, Rett
clinical features, and medication refusal. Definitions are in Methods and the
keyword lists in Supplementary Table~S2.}
\label{fig:pipeline}
\end{figure}

\noindent\textbf{Alt text:} Four-panel study-design figure. Panel A is a
schematic of a five-stage extraction pipeline: keyword filtering of clinical
notes, independent classification by two large language models (a clinical-NLP
expert and a GI/neurology specialist), a consensus gate requiring both to agree,
aggregation into a patient-by-timepoint matrix, and evaluation against ground
truth. Panel B is an icon array of the cohort showing the 115 patients screened, of
whom 108 had QC-confirmed sustained exposure and 56 form the expert-annotated
diarrhea ground truth. Panel C is a donut chart of retrieved-corpus
composition by note type, with telephone encounters the largest slice at 22.7
percent.
Panel D lists the 18 extracted phenotypes grouped into 4 adverse effects, 13
Rett clinical features, and medication refusal.

\begin{figure}[htbp]
\centering
\includegraphics[width=\textwidth]{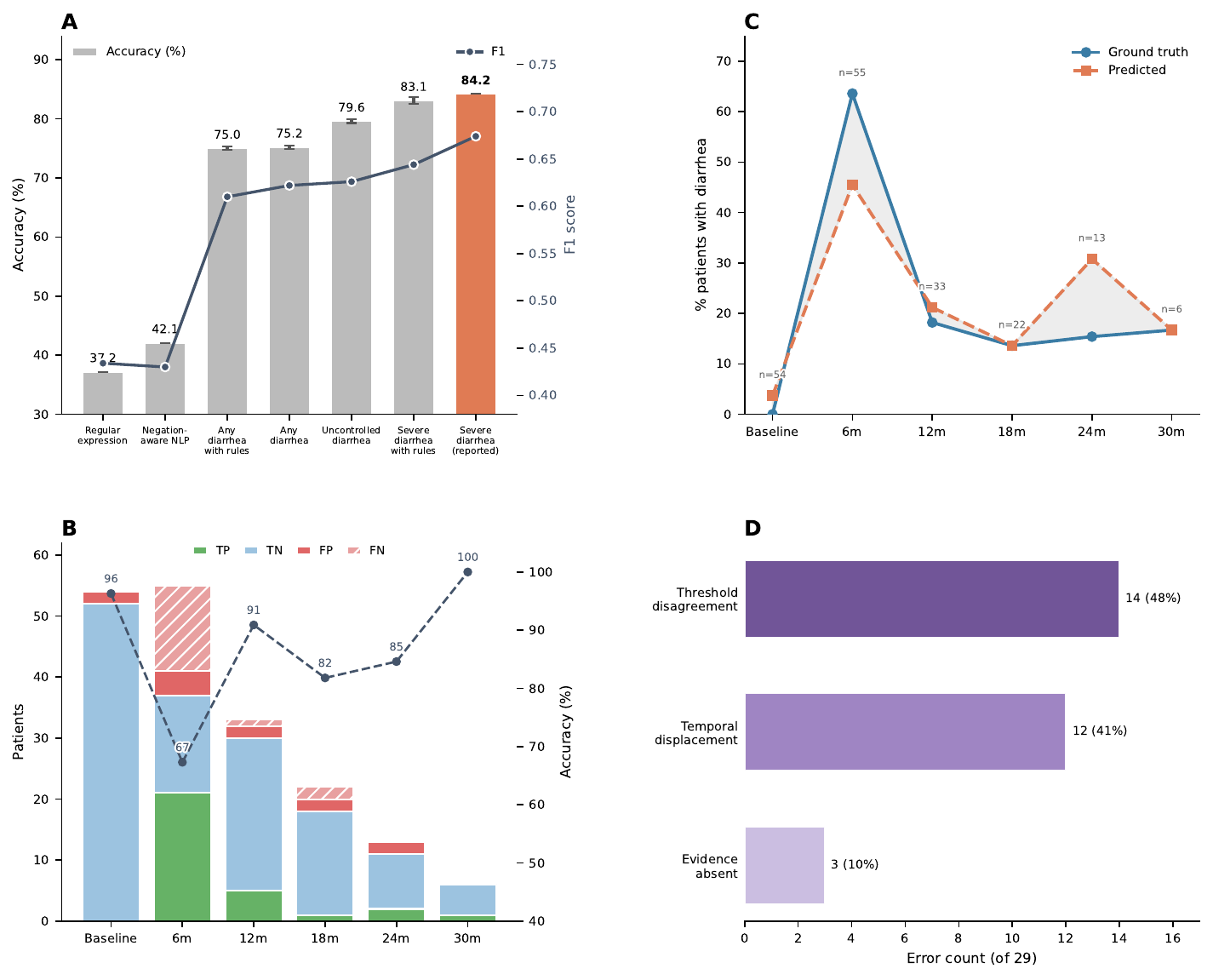}
\caption{Diarrhea validation against the expert ground truth, 183 evaluable
patient-timepoint cells from the 56 labeled patients, all panels on Claude~Opus~4.6.
\textbf{(A)}~Accuracy (bars, error bars = SD over three runs) and F1 (line) for
the seven methods of Table~\ref{tab:model_comparison}, ordered by accuracy, with
the two non-LLM methods shaded apart from the five prompts.
\textbf{(B)}~Confusion structure at each timepoint, stacked true and false
positives and negatives, with overall accuracy as a dashed line. A patient
contributes at most one cell per timepoint, so each bar counts patients.
\textbf{(C)}~Expert and predicted diarrhea prevalence at each timepoint, the
shaded band showing the difference and $n$ the evaluable patients.
\textbf{(D)}~The 29 errors of the reported configuration grouped by mechanism,
with the mechanisms defined in Table~\ref{tab:error_types}.
Panels~B to~D use the reported configuration, whose three runs are identical on
every cell. Statistical tests for panel~A are in Supplementary Table~S4.}
\label{fig:diarrhea}
\end{figure}

\noindent\textbf{Alt text:} Four-panel diarrhea-validation figure. Panel A is a
bar-and-line chart of accuracy and F1 across seven methods, rising from a
regular-expression baseline (37.2 percent) through negation-aware NLP (42.1 percent)
and the two prompts counting any documented diarrhea near 75 percent to the
deployed prompt (79.6 percent) and the two prompts with a severity
bar (83.1 and 84.2 percent). Panel B is stacked
per-timepoint confusion bars with an accuracy line, 96 percent at baseline, with the
6-month timepoint dominated by false negatives and the 24-month timepoint showing
false positives only. Panel C plots ground-truth versus predicted diarrhea prevalence
over time with a shaded difference band, showing under-detection at 6 months and
over-detection at 24 months. Panel D is an error taxonomy of 29 errors by
mechanism: 48 percent threshold disagreement, 41 percent temporal displacement
into an adjacent window, and 10 percent evidence absent from the available notes.

\begin{figure}[htbp]
\centering
\includegraphics[width=\textwidth]{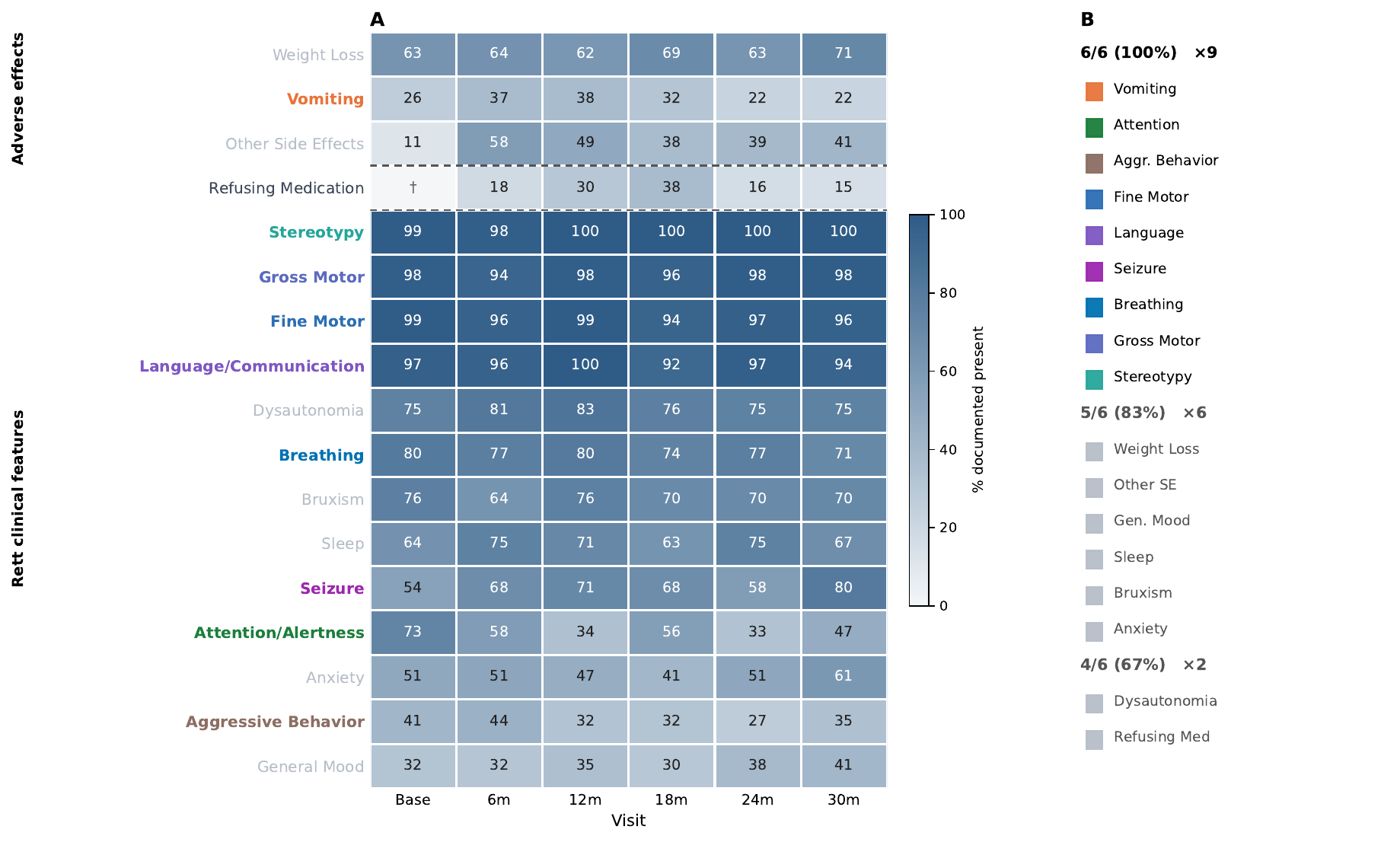}
\caption{Multi-phenotype documented presence across treatment.
\textbf{(A)}~The percentage of patients with each phenotype documented present
at each visit, from Claude~Opus~4.6, on one shared 0--100\% scale. Denominators
are the anchor-confirmed patients with at least one keyword-matching note in that
window, so they vary by phenotype and by visit and are listed in Supplementary
Table~S8. The baseline medication-refusal cell rests on a single patient and is
suppressed ($\dagger$). Dashed lines
separate adverse effects, medication refusal, and Rett clinical features, and that
last block is ordered by mean documented presence. Each visit is a separate cell
and not a point on a connected line, so the panel shows documentation rates rather
than a trajectory of improvement or worsening, and the Rett clinical features are
manifestations of the disorder rather than drug-related harm. Diarrhea is validated separately in
Figure~\ref{fig:diarrhea} and is not plotted here.
\textbf{(B)}~Expert spot-check agreement for the same 17 phenotypes, six cases
each, three model-positive and three model-negative. Row labels in panel~A are
bold for the nine phenotypes reaching 6/6. Full per-visit rates are in
Supplementary Table~S8.}
\label{fig:multipheno}
\end{figure}

\noindent\textbf{Alt text:} Two-panel multi-phenotype documented-presence
figure spanning a pre-treatment baseline through 30 months. Panel A is a
single heatmap of 17 phenotypes by visit, grouped by dashed lines into
adverse effects, medication refusal, and Rett clinical features, the last
being manifestations of the disorder itself rather than adverse effects.
Fine motor, stereotypy, gross motor, and language and communication are
documented in nearly all patients throughout. General mood and aggressive
behavior are documented least often. Weight-loss documentation is high at
baseline and stays high across follow-up, while medication refusal and other
side effects are documented rarely at baseline and rise afterwards. Panel B
summarizes expert spot-check agreement tiers for the 17 non-validated
phenotypes, with nine at full 6/6 agreement shown in colour, six at 5/6, and
two at 4/6 shown in grey.

\end{document}


\begin{center}
{\large\bfseries Supplementary Material}\\[6pt]
{\normalsize\itshape for}\\[2pt]
{\normalsize\bfseries Documenting Trofinetide Treatment Phenotypes in Rett
Syndrome with a Dual-LLM Consensus Pipeline}
\end{center}
\bigskip

\renewcommand{\thetable}{S\arabic{table}}
\renewcommand{\theHtable}{supp.\arabic{table}}
\setcounter{table}{0}

\section*{Supplementary Notes}

\subsection*{Supplementary Note S1. Inter-LLM Agreement and the Consensus Gate}
\label{sec:interllm}

Because only diarrhea has an expert ground truth, we also measured the agreement
between the two independently prompted agents, which requires no labels. Cohen's
$\kappa$ between LLM~1 (clinical NLP expert) and LLM~2 (GI/neurology specialist)
was computed over the 60,530 note--phenotype instances that both models
classified (Figure~\ref{fig:agreement}A). Overall agreement was high
($\kappa=0.930$, raw agreement 96.6\%), and all 17 phenotypes reached
$\kappa\geq0.73$, with per-phenotype values in Table~\ref{tab:kappa_supp}. Seven
of the 60,537 matches did not return a valid YES or NO from both models and were
scored negative. Because the two prompts were written independently, their
agreement is a proxy for how well specified a phenotype is. It does not track the
per-phenotype expert agreement from the spot-check, however. Attention/alertness has
the lowest $\kappa$ (0.732) yet reached 6/6 on review, while dysautonomia
($\kappa=0.917$) and refusing medication ($\kappa=0.890$) reached only 4/6. With six
cases per phenotype the spot-check cannot resolve differences of that size, so the two
signals should be read as independent rather than concordant.

Diarrhea is scored by the diarrhea-specific pipeline rather than by the presence
prompt, and reaches $\kappa=0.878$ at raw agreement 98.0\% over the 2,260 notes both
agents classified. It is therefore shown separately from the 17, as a hatched bar in
Figure~\ref{fig:agreement}A and below the rule in Table~\ref{tab:kappa_supp}.

\paragraph{Prompt wording drives agreement.}
The two diarrhea prompts isolate what wording alone contributes, because both
were applied to the identical 2,260 notes with the same two models. The
deployed prompt~V0 reaches raw agreement 95.1\% and $\kappa=0.805$, and the
reported prompt~V2 reaches 98.0\% and $\kappa=0.878$
(Table~\ref{tab:llm_contingency}). Specifying the target more tightly leaves the
two agents less room to differ, and the same effect appears at the cell level
across the prompt ladder (Table~\ref{tab:prompt_variants}). Among the 17
phenotypes the two lowest values are attention/alertness ($\kappa=0.732$) and the
open-ended other-side-effects category ($0.756$), which are also the two whose
target is least concretely specifiable. Inter-LLM disagreement may therefore be a
useful label-free signal for which phenotypes most need prompt refinement or human
review, although the present spot-check is too small to validate it.

\paragraph{What the consensus gate costs and buys.}
Across the full corpus the two agents disagreed on 3.4\% of instances
(Figure~\ref{fig:agreement}B, with raw counts in
Table~\ref{tab:llm_contingency}). The gate resolves these disagreements
conservatively, counting only ``both~YES'' instances (38.4\%) as positive and
treating one-sided calls as negative. This is how it suppresses one-sided false
positives, at some cost in recall. On the diarrhea ground truth the mechanism is
measurable but small. Under the deployed prompt, the only configuration for which
both agents were also scored alone, the gate changes two of 183 cells against the
stronger of them (Table~\ref{tab:prompt_variants}, panel~A), so it neither costs
nor buys measurable accuracy there. The case for two agents rests on making their
disagreement observable across the 17 phenotypes that have no ground truth.

\paragraph{The voting pattern as a confidence signal.}
Partitioning cells by the two agents' votes shows that cells where both agree No
are the most accurate against the expert labels (84.3\%, 113/134), cells where
both agree Yes are the least accurate (52.3\%, 23/44), and split votes fall
between them (72.7\%, 8/11). The low accuracy of ``both~Yes'' cells is the
threshold-disagreement effect described in the main-text error analysis. The
agents confidently and defensibly detect documented diarrhea that the reviewer
scored as sub-threshold, so positive calls are exactly the cells a deployment
should route for human adjudication.

\paragraph{Why the reported configuration's run-to-run SD is zero.}
Prompt~V2's accuracy in main-text Table~2 carries an SD of 0.0 because its
confusion matrix is identical in all three runs, at 30~true positives,
124~true negatives, 12~false positives and 17~false negatives. This is a property
of the aggregation step and not evidence that the prompt is deterministic. At the
note level the consensus call does move, flipping on 7,~9 and~8 of the 2,260
classified notes across the three run pairs, and on 3,~4 and~3 of the 1,404
notes that fall inside the annotated patients' evaluation windows. None of those
flips changes a cell. Aggregation marks a patient--timepoint \emph{Yes} if
\emph{any} note in the six-month bucket receives a positive call, so a flip on a
note whose bucket already holds another positive is absorbed. The deployed
prompt~V0 flips on 15,~11 and~10 notes over the same pairs, and one of its cells
does move. The zero is therefore correct arithmetic at the cell level while the
underlying note-level classification remains stochastic, and it should not be
read as prompt stability.

\paragraph{Decoding parameters and the independence of the runs.}
No seed and no sampling parameters were set in any of the 42 model-invocation
nodes, so temperature, top-$p$, top-$k$ and the token limit took platform
defaults throughout, as the main-text Methods states. The three runs are
therefore independent draws rather than replays of a single run. We verified this
on the response strings themselves, since the platform exposes no per-call
telemetry. Under V2, Agent~1 returned byte-identical text on 2,243 of the
2,260 notes and Agent~2 on 2,251, whereas a cached or memoized path would
return identical strings everywhere. Byte differences outnumber label flips, so
part of the run-to-run variation is wording that leaves the YES/NO verdict
unchanged.

\subsection*{Supplementary Note S2. Model Development History}
\label{sec:model_history}

This note expands the model-development summary in the main-text Methods. The
pipeline was reached through a sequence of designs, each built to address a
failure mode observed in the previous one, and the benchmark of the six retained
methods against the 189-cell diarrhea ground truth is in
Table~\ref{tab:architecture}.

Two points bound how those numbers relate to the main text. First, the
architecture was selected on the original note corpus, the order-date exposure
anchor and a 189-cell grid, and it is reported here on the numbers it was decided
with. The architecture it settled on was then held fixed, and the prompt
comparison in the main text varies the prompt instead, on the refreshed corpus and
the corrected anchor. The two sets of numbers are not interchangeable. Second, the
\emph{negation-aware NLP} row of main-text Table~2 is not the \emph{rule-based
NLP} of Table~\ref{tab:architecture}. It is a reimplementation written for the
rerun so that every method in Table~2 is scored on one grid, and it reaches
42.1\% there against the 70.9\% of rule-based NLP. The \emph{regular expression} row
is the Stage~0 floor below, and matches its 42.3\% on the original grid.

\paragraph{Stage 0: Regex-only floor.}
The earliest approach flagged any note containing a diarrhea keyword. Recall is
high but precision is poor, because negated, historical, templated and
anticipatory mentions are all counted positive, as in ``denies diarrhea,'' ``had
diarrhea last year,'' checkbox review-of-systems fields and ``call if diarrhea
develops.'' Every subsequent design targets one of these false-positive sources.

\paragraph{Stage 1: Rule-based NLP.}
We added regular-expression negation detection and temporal cue matching to
discount past-tense and resolved mentions. This substantially improved precision
over the floor but was brittle. Hand-written patterns missed paraphrases and
complex sentence structure, and over-aggressive negation occasionally suppressed
true positives.

\paragraph{Stage 2: Hybrid NLP.}
We layered three further components onto the rule-based system. Clinical-form and
template filters remove boilerplate and blank-field matches, severity gates
down-weight ``mild,'' ``controlled'' and ``resolving'' language, and simple
cross-temporal handling was added. At equal overall accuracy to the rule-based model, this
recovered recall that the stricter negation rules had removed, so the two systems
sat at different points on the same precision--recall frontier rather than one
dominating the other.

\paragraph{Stage 3: Iterative single-LLM prototype.}
Before the full ground truth existed, we prototyped a three-pass LLM approach on a
32-patient subset to test whether contextual reasoning could move beyond the
rule-based frontier. It established feasibility but used a different evaluation
set with no paired ground truth, so it is not part of the six-method benchmark.

\paragraph{Stage 4: Single-LLM and the dual-LLM consensus gate.}
Once the full ground truth was available, we evaluated a single
clinical-NLP-expert LLM, whose accuracy (72.7\%) edged above the rule-based
frontier (70.9\%) although the difference was not resolved at 189 cells. Adding a
second, differently prompted domain-specialist model raised the question of how to
combine the two outputs. We tested both aggregation rules on identical model
outputs. An OR-gate (\emph{permissive}, positive if either agent says Yes)
maximizes sensitivity but readmits false positives, while an AND-gate
(\emph{consensus}, positive only if both agree) favors precision. The consensus
gate gave the best overall accuracy and the most run-stable operating point, and
was adopted.

\paragraph{Rejected variant: a broader definition of the target.}
Having fixed the gate, we tried loosening what the prompt counts as a positive.
Both agents were instructed to score any documented diarrhea as present,
dropping the exclusion for diarrhea that was mild or already controlled with
medication, on the reasoning that a symptom under treatment is still a symptom.
This raised recall as intended but added far more false positives than it
recovered true ones, so overall accuracy fell well below the adopted design and
the change was reverted. Note~S7 quantifies the same effect on the reported grid
as versions~V1 and~V4.

\paragraph{What the pairwise tests do and do not establish.}
Table~\ref{tab:architecture}, panel~B, reports all 15 pairwise McNemar exact tests
on the identical 189-cell paired ground truth. Every context-aware method
significantly outperforms the regex floor ($p<0.0001$, and after Holm correction).
The comparisons above the floor are directionally consistent with the accuracy
ordering, with the consensus model winning each head-to-head disagreement, for
example 24 discordant cells to 14 against hybrid NLP, but none reaches
significance ($p=0.11$--$1.0$). This reflects limited power rather than an absence
of effect. The consensus-versus-NLP gap rests on a net of roughly ten cells of
189, and confirming a difference of that size at 80\% power would require on the
order of 550 annotated cells, about three times the current ground truth. The
final rerun sharpens the same reading. On the refreshed corpus and the corrected
anchor the gate reaches 79.6\% and the stronger single agent 79.2\%, differing on
two of 183 cells (Table~\ref{tab:prompt_variants}, panel~A). We therefore adopt
the consensus gate as the best-performing configuration on present evidence, and
for the observability of the two agents' disagreement across the 17 phenotypes
that have no ground truth, rather than as a proven necessity.

\paragraph{Multi-phenotype run configuration.}
The 17-phenotype presence results in main-text Figure~3 and in
Tables~\ref{tab:kappa_supp}, \ref{tab:side_effects}, \ref{tab:improvements} and
\ref{tab:llm_contingency} were produced by the same model as the reported diarrhea
configuration. Both agents ran Claude~Opus~4.6 and issued 60,537 calls each, for
121,074 in total, with temperature, top-$p$, top-$k$, maximum tokens and seed all
left at platform defaults. The denominator for these percentages is the 99
patients whose exposure anchor could be confirmed, which is the 108 of Table~1 less
the 9 who have no reliable start date and therefore no windows to score. The two
paths assign notes to windows slightly differently. The multi-phenotype path
converts a day offset to months by dividing by 30.44 while the diarrhea grid uses
day bounds, which places 26,773 notes in window on the multi-phenotype side
against the 26,763 the diarrhea grid counts. We report the diarrhea grid figure
throughout, since that is the corpus the validated result rests on.

\paragraph{In-sample versus out-of-sample evidence.}
Architecture selection was informed by errors on the same 189 cells used to report
the architecture comparison, and prompt selection by errors on the same 183 cells
used to report the prompt comparison, so both diarrhea accuracies are in-sample
estimates and are upper bounds rather than held-out validations. The architecture, the
consensus gate and the prompt design were then frozen and carried to the remaining 17
phenotypes with only the phenotype definition substituted (Note~S3.4), and their
expert spot-check agreement in the main-text Results is the out-of-sample evidence
that the design transfers.

\subsection*{Supplementary Note S3. LLM Prompt Templates}

Both agents return a structured JSON object rather than a bare token. The output
schema is configured in Palantir Foundry's structured-output interface and appended
to the prompt by the platform rather than typed by the prompt author, and it is
shown here as it is sent to the model. The \texttt{date} field echoes the note's
contact date, which is already known from the EHR metadata before classification
and is bundled into the same record for traceability. It is not an independent date
extracted from the note text. The listings below therefore end with the author's
own instruction to answer YES or NO, and the schema is what the platform appends
after it.

\begin{quote}\small\ttfamily
Output format (JSON): \{"answer": "YES" \textbar\ "NO", "evidence":
"<supporting quote or reasoning>", "date": "<note contact date>"\}
\end{quote}

The two diarrhea prompts are reproduced in full below, as the base-case
version~V0 that Note~S7 sweeps against. Each agent receives a system prompt and a
user prompt. The three independent runs use byte-identical prompt text, so the
run-to-run variation documented in Note~S1 comes from the model rather than from
the prompt, and these listings therefore cover all six V0 classification nodes,
three runs for each of the two agents. \texttt{[prompt\_bucket]} is the only
interpolation point in the system prompts and takes the visit label in words,
meaning Baseline, 6~months, 12~months, 18~months, 24~months or 30~months.
\texttt{[note\_text]} is the note body.

\subsubsection*{S3.1. LLM~1: Clinical NLP Expert, diarrhea prompt V0}

\noindent\emph{System prompt.}
\begin{quote}\small\ttfamily
You are a clinical NLP expert analyzing patient notes for medication side
effects. Determine if the patient is CURRENTLY experiencing clinically
significant diarrhea as a side effect of Trofinetide/Daybue AT THE TIME of this
clinical note.\\[4pt]
RULES:\\
1. ACTIVE current diarrhea at this visit = YES\\
2. Historical references (had diarrhea early on, developed diarrhea when
started) where it has since resolved = NO\\
3. Mild/controlled diarrhea not requiring intervention = NO\\
4. Pre-printed form templates or checkbox fields = NO\\
5. Anticipatory guidance (may cause diarrhea) = NO\\
6. Negated mentions (denies diarrhea, no diarrhea) = NO\\
7. Diarrhea attributed to other causes (URI, illness) = NO\\[4pt]
The note is from approximately [prompt\_bucket] after starting
Trofinetide/Daybue.\\[4pt]
Respond with ONLY: YES or NO
\end{quote}

\noindent\emph{User prompt.}
\begin{quote}\small\ttfamily
Clinical note:\\[4pt]
[note\_text]\\[4pt]
Is the patient CURRENTLY experiencing clinically significant diarrhea as a
medication side effect? YES or NO only.
\end{quote}

\subsubsection*{S3.2. LLM~2: Gastroenterology Specialist, diarrhea prompt V0}

\noindent\emph{System prompt.}
\begin{quote}\small\ttfamily
You are a gastroenterology specialist reviewing clinical notes for active
diarrhea as a side effect of Trofinetide (Daybue) for Rett Syndrome.\\[4pt]
Critical distinctions:\\
- ACTIVE diarrhea NOW causing clinical concern = YES\\
- HISTORICAL diarrhea (had, developed, past tense about dose adjustments) = NO\\
- MILD/CONTROLLED (mostly under control, manageable, only sometimes) = NO\\
- NEGATED (denies, no diarrhea, negative for) = NO\\
- FORM TEMPLATES (checkboxes, blank fields with Severe Diarrhea) = NO\\
- Never has diarrhea = NO\\[4pt]
This note is from the [prompt\_bucket] post-medication time point.\\[4pt]
Respond ONLY: YES or NO
\end{quote}

\noindent\emph{User prompt.}
\begin{quote}\small\ttfamily
Note:\\[4pt]
[note\_text]\\[4pt]
Does this patient have CURRENT clinically significant diarrhea from Trofinetide
at this visit? YES or NO.
\end{quote}

\subsubsection*{S3.3. The reported diarrhea prompt (V2)}

The diarrhea results in the main text use version~V2, which differs from the V0
text above in exactly one rule per agent and is otherwise identical character for
character, including the two role descriptions, the remaining rules, the
visit-label statement, both user prompts and the output schema. The replaced line
is rule~3 in LLM~1 and the \texttt{MILD/CONTROLLED} bullet in LLM~2. The inserted
text below is the same for both agents. The V0-to-V2 contrast is therefore a
single-factor change.

\begin{quote}\small\ttfamily
Diarrhea that is mild, controlled, or adequately managed (responding to
loperamide, Imodium, probiotics, fiber, diet change, or dose adjustment) = NO.
Answer YES only if it is active AND clinically significant, meaning it is
causing ongoing concern or prompting a change in management at this visit.
\end{quote}

V0 states the same exclusion in a single clause, rule~3 above, that mild or
controlled diarrhea not requiring intervention is absent. V2 names the
interventions that count as control and adds the severity bar. That one change is
what separates the two configurations in main-text Table~2, and it is worth
4.6~percentage points against the expert annotation. The difference does not
survive correction for multiplicity, so it is the ranking of a sweep rather than a
demonstrated improvement. The text above is reproduced character for character
from the block that defines V2.

\subsubsection*{S3.4. The presence prompts for the other 17 phenotypes}

The 17 phenotypes that have no ground truth are scored by a separate pair of
presence prompts rather than by the diarrhea prompts above. Those prompts keep the
same two-role design, the same decision rules for historical, negated,
anticipatory and template mentions, and the same interpolated visit label, and
they add a written definition of the phenotype being scored. That definition block
is why the presence prompts carry no diarrhea branch, and why diarrhea is scored
by its own prompt instead (Note~S1). Their full text is in the released code.

\subsection*{Supplementary Note S4. Reading the Error Taxonomy, and Intended Use}
\label{sec:extended_discussion}

\paragraph{How to read the error taxonomy.}
The taxonomy in main-text Table~4 covers the 29 errors of the reported
configuration. Two points about it are worth stating.

The taxonomy does not contain a category for the model misreading a note, and that
absence is a limit of the method rather than a finding. Each error was assigned by
rule, from where the supporting evidence sits relative to the scored window, so the
categories can separate a window-placement mismatch from a threshold disagreement
but cannot tell whether a note was misread within its own window. The prompts emit
only YES or NO, so nothing in the output records the reasoning behind a call.

The 91.8\% quoted in the main text is a bound and not a corrected accuracy. It is
what accuracy becomes if every threshold-disagreement cell is re-scored as correct
wherever the model and the note agree the symptom is present. Reaching it
defensibly would require re-adjudicating the ground truth prospectively, under a
severity threshold written into both the annotation guideline and the prompt,
rather than relabeling the current errors after the fact.

\paragraph{Intended use.}
The operating point we report, precision~0.714 and recall~0.638 on diarrhea, is
well suited to \emph{cohort-level} inference and triage but not to autonomous
per-patient decision-making. At this precision and recall the pipeline is best used
to estimate phenotype prevalence and trajectories across a population, where over-
and under-calls partially offset in aggregate, and to surface candidate notes for a
human reviewer. Errors are also non-random, concentrating in threshold-ambiguous
cases, so any per-patient use should keep a clinician in the loop and treat a
positive call as a prompt for review rather than a conclusion. We therefore frame
the extracted rates as a real-world evidence and hypothesis-generation tool, and
the individual note classifications as a triage aid.

\subsection*{Supplementary Note S5. TRIPOD-LLM Reporting Checklist (Addressed Items)}
\label{sec:tripod_checklist}

We report our study against the TRIPOD-LLM reporting
guideline~\cite{gallifant2025tripod}. The table below lists the checklist items
this manuscript addresses and where each is reported, with items marked
\emph{Partial} addressed in part and the missing component noted. Items not listed
are not applicable to a study that applies an existing, externally developed LLM
through prompting, such as those specific to \emph{de novo} model development,
instruction tuning or model updating. This is not a complete item-by-item
attestation of all 19 items and 50 subitems.

\begin{longtable}{p{0.9cm} p{6.6cm} p{5.4cm}}
\toprule
\textbf{Item} & \textbf{TRIPOD-LLM description (abbrev.)} &
\textbf{Where addressed} \\
\midrule
\endfirsthead
\toprule
\textbf{Item} & \textbf{TRIPOD-LLM description (abbrev.)} &
\textbf{Where addressed} \\
\midrule
\endhead
\bottomrule
\endlastfoot
1   & Title identifies the study as evaluating an LLM, the task, population, and outcome & Title \\
2   & Abstract & Abstract \\
3a  & Healthcare context/use case and rationale & Introduction \\
3b  & Target population and intended use in the care pathway & Introduction; Methods (Study design) \\
4   & Study objectives (development/validation stage) & Introduction \\
5a  & Sources of data and rationale & Methods (Data sources) \\
5b  & Relevant data points and dataset description & Methods (Data sources); Table~1 \\
5c  & Date of oldest and newest text used & Methods (Data sources); analyzed corpus July 2021 to August 2026, retrieved corpus November 1998 to August 2026 \\
5d  & Data preprocessing and quality checking & Methods (Study design, QC review; Pipeline architecture, Steps~1--2) \\
5e  & Handling of missing/excluded data & Methods (Study design, 7 unconfirmed excluded; Pipeline architecture, 9 of 108 without a reliable exposure anchor excluded from the timepoint analysis; Model development, Deceased/NA excluded) \\
6a  & LLM name, version, and training date & Methods (Pipeline architecture; Claude~Opus~4.6, July 2026); Note~S7 (build versions v1.154.0 and v1.151.0) \\
6c  & How text was generated: prompting and inference settings & Methods (Pipeline architecture); Supplementary Note~S3 \\
6d  & Initial and postprocessed output of the LLM & Methods (Pipeline architecture; structured YES/NO output, parsing) \\
6e  & Classification rationale and thresholds & Methods (Pipeline architecture, consensus gate); Supplementary Note~S1 \\
7a  & Metrics capturing output quality vs.\ gold standard & Results (performance); Table~2 \\
7c  & Outcome definition and how predictions were computed & Methods (Pipeline architecture; Model development) \\
7d  & Subjective assessment: assessor qualifications, instructions, agreement & Methods (Model development); Supplementary Note~S6 (qualifications and consensus adjudication, no agreement statistic) \\
7e  & Comparison to other benchmarks/methods & Results (performance, prompt comparison); Supplementary Notes~S2 and~S7, Supplementary Tables~S10 and~S11 \\
8a  & Annotation guidelines with examples & \emph{Partial}; Supplementary Note~S6 (operative definition and severity scale, no written guideline issued) \\
8b  & Number of annotators, double-annotation, interannotator agreement & \emph{Partial}$^{a}$; Supplementary Note~S6 (four reviewers, consensus adjudication, no agreement statistic) \\
8c  & Annotator background/experience & Supplementary Note~S6 (Rett Center clinical team: two pediatric Rett-clinic nurses and two pediatric neurologists) \\
9a  & Prompt design, curation, and selection & Methods (Pipeline architecture; Phenotype definitions); Supplementary Notes~S3 and~S7 \\
9b  & Data used to develop the prompts & Methods (Phenotype definitions); Supplementary Note~S2 \\
12  & Compute/resources required & Discussion; 121,074 agent calls for the multi-phenotype pipeline (60,537 matches $\times$ 2 agents) and 13,560 for the reported diarrhea rerun (2,260 notes $\times$ 2 agents $\times$ 3 runs) \\
13  & Ethics approval and consent/waiver & Methods (Ethical considerations; IRB H-52222) \\
14a & Funding source and role of funders & Funding \\
14b & Conflicts of interest & Conflicts of Interest \\
14e & Availability of study data & Data Availability \\
14f & Availability of code & Data Availability (GitHub) \\
16a & Flow of text/patient data; counts with/without outcome; follow-up & Methods (Study design; Data sources); Figure~1B \\
16b & Participant characteristics overall and by split & \emph{Partial}; Table~1 (\emph{MECP2}/severity omitted) \\
16c & Comparison of key clinical variables across splits & Table~1 \\
16d & Number of participants and events per analysis & Methods (Evaluation; 56 labeled patients, 183 evaluable cells, 47 positive) \\
17  & Performance per prespecified metrics & Results (performance); Supplementary Table~S6 \\
19a & Overall interpretation & Discussion \\
19b & Limitations, biases, statistical uncertainty, generalizability & Discussion; Results (performance; McNemar, CIs) \\
19c & Known data challenges: representation, missingness, bias & Discussion; Methods (Data sources) \\
19d & Intended use, input, and level of autonomy/oversight & Supplementary Note~S4 (Intended use) \\
19e & Handling of poor-quality/unavailable input in practice & Supplementary Note~S4; Discussion \\
19g & Next steps for future research and applicability & Discussion \\
\end{longtable}

\noindent$^{a}$\textbf{Item 8b (inter-annotator agreement).} The ground truth was
set by the Rett Center clinical team of two pediatric Rett-clinic nurses and two
pediatric neurologists, who resolved each cell together by consensus rather than
annotating independently, so no $\kappa$ or F1 agreement statistic can be computed
retrospectively (Note~S6). This is a real limitation for gold-standard reliability
and is part of why reviewer--model threshold disagreements dominate the error
analysis.

\subsection*{Supplementary Note S6. Annotation Procedure}
\label{sec:annotation_guideline}

The diarrhea ground truth and the multi-phenotype expert spot-check were
established by the Rett Center clinical team, comprising two pediatric nurses
familiar with the Rett syndrome clinic and two pediatric neurologists, who worked
as a single reviewing group throughout. For each patient and timepoint window the
reviewers examined the available clinical notes and labeled the cell Yes for
medication-associated diarrhea, No, or Deceased/Not~applicable. Where diarrhea was
recorded, its severity was noted on the clinic's four-level scale: 1, loose stool in
diaper; 2, watery stool in diaper; 3, watery stool on clothes; 4, watery stool
outside clothes. The operative question was
whether documented diarrhea represented a medication side effect as opposed to a
tolerated, controlled or pre-existing baseline symptom. This is the same judgment
that drives the reviewer--model threshold disagreements in main-text Table~4.

The reviewers resolved each cell together by consensus rather than annotating
independently against a written guideline. No formal written guideline was issued,
and because each cell was adjudicated by consensus, no inter-annotator agreement
statistic could be computed retrospectively. This is a genuine limitation for
gold-standard reliability (TRIPOD-LLM items 8a--8c). A shared written guideline with
an explicit severity threshold, applied by independent double annotation, would both
tighten the ground truth and allow the reviewer--model threshold disagreements to be
quantified rather than adjudicated case by case.

\paragraph{Scope of the spot-check sample.}
The spot-check sample was drawn before the exposure-anchor adjudication was final,
so its coverage of the reported evaluation windows is partial. Of the 102 reviewed
notes, 58 fall inside an evaluation window in the reported configuration and 44 do
not. The shortfall is concentrated at baseline, where 26 of 33 cases fall outside,
against 12 of 34 at 6~months, 6 of 34 at 12~months and 0 of 1 at 18~months. That
gradient follows from the baseline window being the one defined immediately around
the anchor, so it is the most sensitive to where the anchor sits, while the later
windows are progressively less so. A failure of record linkage would instead have
removed cases uniformly. The shortfall is balanced with respect to the model's call,
affecting 22 of the 51 model-positive and 22 of the 51 model-negative cases, so the
58 are not tilted toward agreement or disagreement. On those 58 the reported
configuration returns the same call in 57 cases, and expert agreement over them is
91.4\% (53/58). The 90.2\% (92/102) in the main text is the figure the reviewers
produced over the notes they read, and we report it as such. Re-estimating agreement
directly on the reported configuration would require a fresh sample.

\clearpage
\subsection*{Supplementary Note S7. Prompt-Level and Variable-Decomposition
Sensitivity Analyses}

Note~S2 describes how the \emph{architecture} was selected. Holding that
architecture fixed, we ran three further experiments on the refreshed corpus and
the corrected exposure anchor to ask which component of the configuration the
remaining error is attributable to. The first varied the prompt wording, and it is
where the reported configuration comes from. The second replaced the single
extraction question with two simpler questions combined by an explicit rule. The
third left the extraction untouched and realigned each patient's sequence of
predictions in time. The main text carries the prompt comparison as Table~2, while
the realignment appears only here, because it presumes that onset falls at the
first on-treatment visit and is therefore a diagnostic of where the error sits
rather than a deployable comparator.

The version identifiers used in this supplement and in the released code
correspond to the main-text names as follows. V0 is \emph{uncontrolled diarrhea},
the deployed prompt; V1 is \emph{any diarrhea}; V2 is \emph{severe diarrhea}, the
reported prompt; V3 is \emph{severe diarrhea with rules}; V4 is \emph{any diarrhea
with rules}; and S1 is the \emph{onset-aligned} variant. \emph{Severe} denotes the
clinical-significance bar quoted verbatim in Note~S3, that the diarrhea be active
and either causing ongoing concern or prompting a change in management.
\emph{With rules} denotes the pair of rules that treat resolved diarrhea, and
diarrhea attributed to another cause, as absent.

\subsubsection*{Prompt wording}

Five prompt versions were compared on the same 183 annotated diarrhea cells, with
the two agents, the consensus gate, the keyword pre-filter and the output schema
all unchanged. Two elements were varied. The first is the severity threshold,
specifically whether diarrhea being controlled with medication counts as present.
The second is a pair of explicit reasoning rules stating that a resolution later in
a note governs an earlier onset, and that a finding attributed to another cause is
absent. Version~V0 is the deployed prompt and serves as the base case. Each version
was run three times. Table~\ref{tab:prompt_variants} reports the results, and three
findings follow.

First, \textbf{the severity threshold is the only element that moves accuracy}.
Restating the deployed exclusion explicitly and adding a severity bar raised
accuracy from 79.6\% to 84.2\% (V0 to V2), the largest movement anywhere in the
family. The mechanism is precision. False positives fall from 21 to 12 while true
positives are essentially unchanged at 31 and 30, so the added bar suppresses calls
on documented but managed diarrhea without giving up detections. The difference is
nonetheless not established, as it rests on 9 cells won against 1 lost, which is
$p=0.021$ before correction and $p=0.15$ after Holm correction across the seven
comparisons in this family.

Second, \textbf{the explicit reasoning rules add nothing}. V3, which is V2 plus
those rules, scores 83.1\% against V2's 84.2\%, differing on two cells of 183
($p=0.50$), and V0 to V3 changes 12 cells for a net of six ($p=0.35$ after
correction). The reason is that the deployed prompt already carried both
instructions in weaker form, stating that historical references where the finding
has clearly resolved are absent, and that findings attributed to another cause are
absent (Note~S3). The added rules therefore restated existing guidance rather than
supplying a new criterion. The residual temporal errors in main-text Table~4
reflect inconsistent application of a rule the prompt already contained, and
restating that rule more forcefully did not correct them.

Third, \textbf{reclassifying medication-controlled diarrhea as present made
performance worse}, by 4.4~points in V1 and 4.6~points in V4, consistently across
all three runs. The mechanism is a poor trade, adding 14~false positives to remove
6~false negatives. This is the same direction as the broader definition of the
target that was tested and reverted during development (Note~S2), now quantified on
the reported grid, and it is why the deployed prompt treats adequately controlled
diarrhea as absent. Neither broader
version is significantly worse than V0 on this grid ($p=0.35$ and $p=0.25$ after
correction), so the threshold choice is supported by a consistent direction across
runs and versions rather than by a resolved test.

Model count is not one of the elements that matters here. Running either agent
alone under V0 gives 78.5\% and 79.2\%, and the gate's 79.6\% differs from the
stronger single agent on two cells, so the whole spread in
Table~\ref{tab:prompt_variants} is attributable to wording rather than to the
number of calls (Note~S1).

The ranking also survives dropping the pre-treatment baseline. The 54 baseline cells
are almost all true negatives and inflate every configuration equally, and on the
129~on-treatment cells alone the deployed prompt scores 72.9\%, the reported prompt
79.1\% at precision 0.750, and the onset-aligned variant 79.1\% at recall 0.745.
Removing baseline therefore costs the reported configuration 5.1~points without
changing which configuration wins. These three are computed from the modal
predictions rather than as three-run means, because the per-timepoint confusion
counts were exported for the modal run only, which is also why the two rule-based
methods and versions~V1, V3 and~V4 cannot be shown this way.

\subsubsection*{Decomposing the extraction question}

The prompt experiment leaves open the possibility that the single extraction
question asks for two judgments at once, whether diarrhea has started and whether it
is currently uncontrolled. We tested that reading by replacing the single question
with two simpler ones and combining them in code rather than in the prompt.
Variable~A asks only whether diarrhea has begun as a side effect of trofinetide at
any point up to and including the visit. Variable~B asks only whether it is
adequately controlled at the visit. The two were combined by a deterministic state
rule that labels diarrhea present from the visit at which onset is first detected
and absent from the visit at which control is first established. Ten combination
rules were evaluated, varying that rule, the direction of the consensus gate and the
within-window aggregation order, all at no additional inference cost, since each is
a function of the same two extracted variables.

The decomposition did \textbf{not improve on the single question}. The ten rules
span 61.7\% to 79.4\% (Table~\ref{tab:prompt_variants}, panel~B), the best coming
within 0.2~points of the deployed prompt and none reaching the 84.2\% of the
reported configuration. The onset variable is what limits them. Applied alone it
scores 61.7\% at precision 0.384, so it fires on 61 of the 136~negative cells, and
because the state rule cannot label a visit negative until control is documented,
the combination rules can only recover accuracy by suppressing harder, which
collapses recall. As with the reasoning rules above, the deployed prompt already
encodes this logic, since it scores active diarrhea as present and both resolved
history and adequately controlled disease as absent (Note~S3). Splitting that
judgment into two questions and a state machine reproduced it less accurately than
asking it once.

\subsubsection*{Realigning the predictions in time}

Correcting the exposure anchor removes the dispensing and titration delay from the
window boundaries (Note~S8), but it cannot remove the lag between a symptom starting
and a note recording it. A third experiment asked what accuracy would be available
in the best case, if the anchor and the annotation agreed on when diarrhea began. We
left the extraction untouched and slid each patient's sequence of predicted visits so
that the first visit predicted positive falls at the first on-treatment visit, padding
the vacated tail as negative. Only the model output moves, so every cell remains
comparable against the same annotation. Applied to the deployed prompt's modal
predictions this raised accuracy from 79.8\% to 84.2\%, the best F1 anywhere in the
family at 0.707 (variant~S1 in Table~\ref{tab:prompt_variants}, panel~C), matching
the reported prompt from the opposite direction, by cutting false negatives rather
than false positives.

This is a bound and not a method. Collapsing every patient's onset to the first
on-treatment visit removes any possibility of predicting a later onset, so it assumes
away the case of a patient who develops the symptom after a year. For diarrhea that
assumption is nearly free, since the annotation places 35 of the 37 positive patients
at their first on-treatment visit, but it is not an assumption we could carry to the
other phenotypes, where onset is not concentrated at a single visit. A related
benchmark makes the same point about the task. Predicting from the visit number
alone, positive at 6~months and negative everywhere else, scores 82.5\% on these
cells, because 35 of the 47 positive cells sit at that one visit. Much of the
annotated signal is carried by the timing of the visit relative to treatment start
rather than by the content of the individual note, which is a property of this
label set rather than a result about the model.

\paragraph{How these experiments sit in the development process.}
The results above are the final stage of an iterative process and not a set of
independent replications. We first searched over architectures, then over prompts,
running rounds of error analysis between stages and carrying the surviving
configuration forward each time. Two things changed along the way. Error analysis
showed that some of the remaining error was in the data rather than in the model, so
the note pull was extended from one ending in 2025 to one extending through August
2026, and the exposure anchor was replaced (Note~S8). Earlier stages were therefore
run on a different corpus and a different set of evaluation windows, and their
numbers are not comparable to the ones reported here. Where an earlier stage and the
final run disagree, we report the final run.

Model builds differ between the two for the same reason. Every configuration of the
final rerun ran in one session on Claude~Opus~4.6~v1.154.0 and the arms are therefore
mutually comparable, while the architecture comparison in
Table~\ref{tab:architecture} ran on the build the pipeline was originally deployed
with, v1.151.0. The platform withdrew v1.151.0 before the rerun, so a byte-identical
replication of the deployed configuration was not available and the two sets are
reported separately rather than pooled.

\subsection*{Supplementary Note S8. The Exposure Anchor}

Every follow-up window is measured from the date trofinetide was started, so that
date determines which visit a note is scored against. The deployed pipeline took it
from the earliest trofinetide record in the Epic order table, and the resulting
windows disagreed with the expert annotation more often than the note content did.
Manual review of the record explained why. Trofinetide is an oral solution dispensed
by a specialty pharmacy and given at home, so the drug appears in orders, charts and
correspondence well before the family gives the first dose, and the order is followed
by pharmacy processing, insurance authorization and a four-week titration. In one
reviewed patient the prescription reached the specialty pharmacy 73~days after the
order, insurance declined coverage the same day, and the first record of the family
giving the drug is at day~114. In another the order preceded the first documented
dose by more than two years. In both cases the agents read every note correctly and
the error was in which window the note had been placed.

We therefore replaced the order date with the earliest documented \emph{exposure},
taken in order of preference from the medication administration record, then from a
dispense or fill record, and otherwise from the earliest note adjudicated as
documenting active dosing. Of the 108~QC-confirmed patients, 8 anchor on an
administration record and 91 on an adjudicated note, while 9 have no reliable
evidence of a start date and are excluded from the timepoint analysis. None anchors
on a dispense record, because trofinetide fills go through a specialty pharmacy and
never reach the institutional dispense tables. The correction is large and does not
depend on the labels, moving the median patient by 116~days, with more than half
moving by more than 90~days. It also moves six
cells out of the follow-up range, taking the grid from 189 cells to 183 while
conserving the contributing patients and the 47 positive cells, and accuracy under
the deployed prompt rises from 77.1\% to 79.6\%. That rise cannot be attributed to
the anchor alone, because the note corpus was refreshed in the same rerun, so we
report the two grids side by side. Two negative controls confirm that the result does
not follow from displacing windows as such, since neither randomly re-anchoring the
patients nor permuting their timepoint labels reaches the observed accuracy in 200
draws.

The corrected anchor does not remove the problem entirely. It cannot recover the lag
between a symptom starting and a note recording it, and the annotation is cell-level
and carries no dates, so a reviewer may place a symptom at the stage of treatment
where it clinically belongs rather than at the timestamp of the note that records it.
Both effects are visible in the gap between the two ways of scoring the pipeline.
Judged note by note, where the anchor plays no part, agreement with expert review is
about 90\%, at 90.2\% over the full 17-phenotype spot-check and 91.4\% over the 58
of those cases that fall inside an evaluation window (Note~S6). Judged cell by cell, where
every call must also land in the right six-month window, the reported configuration
reaches 84.2\%. We read the difference as the cost of identifying the start date from
the record rather than as a failure to read the notes, which is why the residual
temporal errors in main-text Table~4 are the largest remaining category after
threshold disagreement.

\paragraph{Denominators in main-text Table~1.}
The 9 patients without a confirmed start date have no evaluation window, so the age,
follow-up, notes-per-patient and total-analyzed-notes rows rest on the remaining 99,
while the total-retrieved-notes and sex rows cover all 108. The comparison group is therefore 43 patients on the
former and 52 on the latter. Follow-up is measured from the anchor to each patient's
last available note rather than to the corpus cutoff, and age is computed from the
date of birth in the institutional EHR demographics table.

\clearpage
\section*{Supplementary Tables}

\setcounter{table}{0}

\begin{table}[H]
\centering
\caption{Note-type distribution of the 91,453 notes retrieved for the 108
QC-confirmed patients, giving the top~10 types by volume with 57 further types
collapsed. Corresponds to main-text Figure~1C. The last two columns give the
diarrhea funnel, first the notes containing at least one term from the diarrhea
keyword list, then the subset of those falling inside a patient's evaluation window
and therefore classified by both agents. Retrieval is broader than analysis, and
the 26,763 notes falling inside an evaluation window are the analyzed subset
counted in main-text Table~1. The 1,650~MR~AVS~Snapshot notes illustrate the
distinction, contributing 157 keyword-passing notes but no classifications, because
none falls inside a window.}
\label{tab:notetype_supp}
\begin{tabular}{lcccc}
\toprule
\textbf{Note Type} & \textbf{Notes} & \textbf{\% Corpus} & \textbf{Keyword-passing} & \textbf{Classified} \\
\midrule
Telephone Encounter    & 20,760 & 22.7\% & 425 & 164 \\
Progress Notes         & 20,683 & 22.6\% & 3,125 & 879 \\
(Unspecified)          & 15,899 & 17.4\% & 381 & 177 \\
Letter                 &  8,584 &  9.4\% & 1,399 & 536 \\
Nsg/Anc Progress Note  &  7,753 &  8.5\% & 86 & 36 \\
Plan of Care           &  2,965 &  3.2\% & 11 & 2 \\
MD Progress Note       &  2,923 &  3.2\% & 173 & 80 \\
MR AVS Snapshot        &  1,650 &  1.8\% & 157 & 0 \\
Patient Instructions   &  1,604 &  1.8\% & 98 & 12 \\
Addendum Note          &  1,491 &  1.6\% & 0 & 0 \\
\midrule
Other (57 types)       &  7,141 &  7.8\% & 1,256 & 374 \\
\midrule
\textbf{Total}         & \textbf{91,453} & \textbf{100\%} & \textbf{7,111} & \textbf{2,260} \\
\bottomrule
\end{tabular}
\end{table}

\begin{table}[H]
\centering
\caption{Representative keywords for each phenotype category. The lists are
abbreviated here, and the full lists contain 14--22 terms per phenotype and 24 for
diarrhea.}
\label{tab:keywords}
\begin{tabular}{ll}
\toprule
\textbf{Phenotype} & \textbf{Representative Keywords} \\
\midrule
Diarrhea     & diarrhea, loose stool, watery stool, pasty stool, GI side effect \\
Vomiting     & vomiting, emesis, retching, gagging, nausea \\
Weight Loss  & weight loss, poor appetite, failure to thrive, eating less \\
Mood         & irritable, low mood, distress, dysphoric, agitated \\
Gross Motor  & gait difficulty, impaired balance, non-ambulatory, unsteady \\
Stereotypy   & hand wringing, hand mouthing, midline hand movements, stereotyped \\
Seizure      & seizure, epileptic event, convulsion, ongoing seizures \\
\bottomrule
\end{tabular}
\end{table}

\begin{table}[H]
\centering
\caption{Inter-LLM agreement between LLM~1 (clinical NLP expert) and LLM~2
(GI/neurology specialist) for each phenotype (Claude~Opus~4.6), given as Cohen's
$\kappa$ and raw agreement over the note--phenotype instances that returned a valid
response from both models. Table~\ref{tab:llm_contingency} gives the raw
contingency counts, and Figure~\ref{fig:agreement} plots these values. Sorted by
$\kappa$. The diarrhea row uses the severe-diarrhea prompt~V2 on its own validated
2,260-note corpus, a different prompt from the other 17, so it is listed separately
and excluded from the overall row.}
\label{tab:kappa_supp}
\begin{tabular}{lccc}
\toprule
\textbf{Phenotype} & \textbf{N instances} & \textbf{Raw agreement} & \textbf{Cohen's $\kappa$} \\
\midrule
\textbf{Overall (17 phenotypes)} & \textbf{60,530} & \textbf{96.6\%} & \textbf{0.930} \\
\midrule
Vomiting                 & 2,959 & 98.3\% & 0.943 \\
Aggressive Behavior      & 4,297 & 98.9\% & 0.932 \\
Gross Motor              & 6,022 & 97.8\% & 0.931 \\
Dysautonomia             & 2,699 & 96.6\% & 0.917 \\
Weight Loss              & 1,905 & 95.6\% & 0.910 \\
Seizure                  &   562 & 96.3\% & 0.908 \\
Breathing                & 6,192 & 96.3\% & 0.905 \\
Language/Communication   & 8,277 & 95.9\% & 0.901 \\
Anxiety                  & 5,299 & 97.6\% & 0.898 \\
Bruxism                  & 1,113 & 96.1\% & 0.897 \\
Stereotypy               & 1,423 & 99.4\% & 0.892 \\
Refusing Medication      &   566 & 97.0\% & 0.890 \\
General Mood             & 5,973 & 98.4\% & 0.884 \\
Sleep                    & 4,884 & 94.3\% & 0.876 \\
Fine Motor               & 3,443 & 95.1\% & 0.826 \\
Other Side Effects       & 4,603 & 94.1\% & 0.756 \\
Attention/Alertness      &   313 & 86.6\% & 0.732 \\
\midrule
Diarrhea (severe, V2)    & 2,260 & 98.0\% & 0.878 \\
\bottomrule
\multicolumn{4}{l}{\small Stereotypy is documented present in 97\% of its classified notes;} \\
\multicolumn{4}{l}{\small that near-universal prevalence caps $\kappa$ despite 99.4\% raw agreement.} \\
\end{tabular}
\end{table}

\begin{table}[H]
\centering
\caption{Pairwise McNemar exact tests among the comparator configurations on the
183-cell diarrhea ground truth, computed on the modal prediction across three runs.
$b$ counts the cells where the first configuration is correct and the second wrong,
$c$ the reverse, and $p$ is the two-sided exact $p$-value. $p_{\text{Holm}}$ is
Holm-adjusted \emph{within} the pre-declared family the comparison belongs to,
which is of size 7 for the prompts, 1 for the alignment contrast and 2 for the
rule-based floors, rather than across all comparisons. Significant results
($p<0.05$) in bold. Only the two floors are separated from the reported prompt. No
contrast within the prompt family clears Holm correction, which is why V2 is
described as the top of a sweep rather than as a demonstrated improvement. Every
test is cell-comparable, meaning both configurations are scored against the same
annotation on the same windows.}
\label{tab:mcnemar}
\small
\begin{tabular}{llcccc}
\toprule
\textbf{Configuration A} & \textbf{Configuration B} & \textbf{$b$} & \textbf{$c$} & \textbf{$p$} & \textbf{$p_{\text{Holm}}$} \\
\midrule
\multicolumn{6}{l}{\textit{Prompts (family of 7)}} \\
V2 & V0 (as deployed)      &  9 &  1 & 0.021 & 0.149 \\
V2 & V3                    &  2 &  0 & 0.500 & 0.500 \\
V2 & Agent~1 alone (V0)    & 13 &  3 & 0.021 & 0.149 \\
V2 & Agent~2 alone (V0)    & 10 &  2 & 0.039 & 0.193 \\
V0 & V1                    & 14 &  6 & 0.115 & 0.346 \\
V0 & V3                    &  3 &  9 & 0.146 & 0.346 \\
V0 & V4                    & 14 &  5 & 0.064 & 0.254 \\
\addlinespace
\multicolumn{6}{l}{\textit{Temporal alignment (family of 1)}} \\
V2 & S1 (V0 realigned)     &  9 &  9 & 1.000 & 1.000 \\
\addlinespace
\multicolumn{6}{l}{\textit{Rule-based floors (family of 2)}} \\
V2 & Regular expression     & 100 & 14 & \textbf{$<$0.0001} & \textbf{$<$0.0001} \\
V2 & Negation-aware NLP     &  88 & 11 & \textbf{$<$0.0001} & \textbf{$<$0.0001} \\
\bottomrule
\end{tabular}
\end{table}

\begin{table}[H]
\centering
\caption{Per-timepoint performance of the reported diarrhea configuration
(Claude~Opus~4.6, prompt~V2, severe diarrhea, under the dual-LLM consensus gate).
All three runs returned identical predictions on all 183 cells, so these are the
predictions themselves rather than a representative run, although the note-level
calls behind them are not fully stable (Note~S1). Baseline is not negative by
construction, since two of the 54 baseline cells are false positives.}
\label{tab:per_timepoint}
\begin{tabular}{lcccccc}
\toprule
\textbf{Timepoint} & \textbf{N} & \textbf{Accuracy} & \textbf{TP} & \textbf{TN} & \textbf{FP} & \textbf{FN} \\
\midrule
Baseline   & 54 & 96.3\%  & 0  & 52 & 2  & 0  \\
6 months   & 55 & 67.3\%  & 21 & 16 & 4  & 14 \\
12 months  & 33 & 90.9\%  & 5  & 25 & 2  & 1  \\
18 months  & 22 & 81.8\%  & 1  & 17 & 2  & 2  \\
24 months  & 13 & 84.6\%  & 2  & 9  & 2  & 0  \\
30 months  & 6  & 100.0\% & 1  & 5  & 0  & 0  \\
\bottomrule
\end{tabular}
\end{table}

\begin{table}[H]
\centering
\caption{Additional per-timepoint metrics for the reported diarrhea configuration
(Claude~Opus~4.6, prompt~V2 under the consensus gate), computed from the same
confusion counts as Table~\ref{tab:per_timepoint}, namely 30~TP, 124~TN, 12~FP and
17~FN over 183 evaluable cells. BalAcc is balanced accuracy, PPV precision, Sens
recall, Spec specificity, NPV negative predictive value, and MCC the Matthews
correlation coefficient. Dashes mark metrics undefined at a timepoint, which at
baseline is every metric requiring a positive gold case. The rows at 24 and
30~months rest on 13 and 6~cells and should be read as descriptive.}
\label{tab:extended_metrics_supp}
\begin{tabular}{lccccccccc}
\toprule
\textbf{Timepoint} & \textbf{Acc} & \textbf{BalAcc} & \textbf{PPV} & \textbf{Sens} & \textbf{Spec} & \textbf{F1} & \textbf{NPV} & \textbf{MCC} & \textbf{Prev} \\
\midrule
Overall   & .842 & .775 & .714 & .638 & .912 & .674 & .879 & .571 & .257 \\
Baseline  & .963 & -- & .000 & -- & .963 & .000 & 1.00 & -- & .000 \\
6 months  & .673 & .700 & .840 & .600 & .800 & .700 & .533 & .386 & .636 \\
12 months & .909 & .880 & .714 & .833 & .926 & .769 & .962 & .716 & .182 \\
18 months & .818 & .614 & .333 & .333 & .895 & .333 & .895 & .228 & .136 \\
24 months & .846 & .909 & .500 & 1.00 & .818 & .667 & 1.00 & .640 & .154 \\
30 months & 1.00 & 1.00 & 1.00 & 1.00 & 1.00 & 1.00 & 1.00 & 1.00 & .167 \\
\bottomrule
\end{tabular}
\end{table}

\begin{table}[H]
\centering
\caption{Adverse effects, given as the percentage of patients with the finding
documented as present at each visit (Claude~Opus~4.6). Base is the pre-treatment
baseline. A patient contributes a cell at a visit only where a keyword-passing note
falls inside that window, so denominators vary by phenotype and visit, and the last
column gives the range of per-visit denominators across the six visits.
$\dagger$~marks a cell suppressed because it rests on a single patient. Diarrhea is
not listed here because it is scored by the severe-diarrhea prompt~V2 against its
own ground truth, with per-visit performance in
Table~\ref{tab:per_timepoint}.}
\label{tab:side_effects}
\begin{tabular}{lccccccc}
\toprule
\textbf{Phenotype} & \textbf{Base} & \textbf{6m} & \textbf{12m} & \textbf{18m} & \textbf{24m} & \textbf{30m} & \textbf{$n$ range} \\
\midrule
Weight Loss             &  63.2 &  64.2 &  61.7 &  69.2 &  62.7 &  71.4 &  42--67 \\
Vomiting                &  26.5 &  37.3 &  37.5 &  31.6 &  22.2 &  21.7 &  46--75 \\
Other Side Effects      &  11.0 &  57.8 &  49.4 &  37.7 &  39.3 &  41.2 &  51--90 \\
Refusing Medication     & $\dagger$ &  18.2 &  30.3 &  37.5 &  16.1 &  14.8 &  24--33 \\
\bottomrule
\end{tabular}
\end{table}

\begin{table}[H]
\centering
\caption{Rett clinical features, given as the percentage of patients with the
finding documented as present at each visit (Claude~Opus~4.6). These features are
manifestations of Rett syndrome itself rather than adverse effects of trofinetide.
Values reflect documented presence of the finding regardless of whether it is
improving or worsening, and Base is the pre-treatment baseline. A patient
contributes a cell at a visit only where a keyword-passing note falls inside that
window, so denominators vary by phenotype and visit, and the last column gives the
range of per-visit denominators across the six visits. Documented presence is lower
at baseline than at 6 months for 5 of the 13 features, so baseline-to-follow-up
differences in this table should not be interpreted as clinical change.}
\label{tab:improvements}
\begin{tabular}{lccccccc}
\toprule
\textbf{Phenotype} & \textbf{Base} & \textbf{6m} & \textbf{12m} & \textbf{18m} & \textbf{24m} & \textbf{30m} & \textbf{$n$ range} \\
\midrule
Attention/Alertness     &  73.3 &  58.1 &  34.4 &  56.5 &  33.3 &  47.4 &  15--32 \\
General Mood            &  32.1 &  31.5 &  35.0 &  30.3 &  38.3 &  40.7 &  54--89 \\
Gross Motor             &  97.6 &  93.5 &  97.6 &  95.7 &  98.4 &  98.1 &  53--93 \\
Fine Motor              &  98.8 &  95.5 &  98.7 &  94.0 &  96.6 &  95.9 &  49--89 \\
Dysautonomia            &  74.7 &  81.2 &  83.3 &  76.1 &  75.0 &  75.0 &  48--85 \\
Language/Communication  &  96.6 &  95.7 & 100.0 &  91.9 &  96.9 &  94.5 &  55--93 \\
Seizure                 &  54.5 &  67.6 &  70.6 &  67.7 &  57.7 &  80.0 &  20--34 \\
Stereotypy              &  98.8 &  97.5 & 100.0 & 100.0 & 100.0 & 100.0 &  45--80 \\
Breathing               &  80.2 &  77.3 &  79.7 &  73.9 &  77.0 &  70.6 &  51--88 \\
Sleep                   &  64.3 &  75.3 &  70.9 &  63.4 &  75.0 &  67.3 &  52--89 \\
Bruxism                 &  76.1 &  63.6 &  75.7 &  69.6 &  69.8 &  70.5 &  44--77 \\
Anxiety                 &  50.7 &  51.2 &  47.2 &  41.0 &  50.9 &  60.9 &  46--82 \\
Aggressive Behavior     &  41.2 &  44.3 &  32.5 &  32.4 &  26.7 &  34.6 &  52--88 \\
\bottomrule
\end{tabular}
\end{table}

\begin{table}[H]
\centering
\caption{Contingency table of LLM~1 (clinical NLP expert) and LLM~2 (domain
specialist) calls (Claude~Opus~4.6). The first row covers the 17 phenotypes scored
by the presence prompt, restricted to the 60,530 of 60,537 note--phenotype matches
that returned a valid response from both models, which is the same base
Table~\ref{tab:kappa_supp} uses. The two diarrhea rows use the diarrhea-specific
prompts on the identical 2,260-note corpus, so the difference between them isolates
the effect of prompt wording on how often the two agents differ. Their cells are
means over the three runs and are therefore not integers.}
\label{tab:llm_contingency}
\begin{tabular}{lccccc}
\toprule
\textbf{Phenotype} & \textbf{Both YES} & \textbf{LLM~1 only} & \textbf{LLM~2 only} & \textbf{Both NO} & \textbf{$N$} \\
\midrule
Presence prompt, 17 phenotypes  & 23,251 & 1,123 & 923 & 35,233 & 60,530 \\
\addlinespace
Diarrhea, V0 (deployed)         &    276 &  97.7 & 12.7 & 1,873.3 & 2,260 \\
Diarrhea, V2 (reported)         &    183 &  25.3 & 20.3 & 2,031.3 & 2,260 \\
\bottomrule
\end{tabular}
\end{table}

\begin{table}[H]
\centering
\small
\caption{Prompt-level, variable-decomposition and temporal-alignment sensitivity
analyses (Note~S7). Every variant is scored on the same 183 annotated diarrhea
cells, with the architecture, consensus gate, keyword pre-filter and output schema
held fixed, so every $\Delta$ is a like-for-like comparison. V0 is the deployed
prompt and serves as the base case, and V2 is the configuration reported in the main
text. Accuracy in panels~A and~B is the mean$\pm$SD of three independent runs, as in
main-text Table~2, and the other metrics are computed from the modal prediction
across those runs. Panel~C is a deterministic transformation of the V0 modal
predictions, so it was run once and its $\Delta$ is taken against those predictions.
The $p$~values are McNemar exact against the base case, Holm-adjusted within the
family the comparison belongs to (Table~\ref{tab:mcnemar}), and are therefore not
comparable across panels. Three readings follow. The severity threshold in panel~A
is the only element that moves accuracy materially. No decomposition rule in
panel~B beats the deployed prompt, and the best comes within 0.2~points of it. In
panel~C, S1 shifts only the model output and is therefore scored against the same
annotation on the same cells, whereas S3 shifts the annotation as well and is
positional, so its 85.2\% is an upper bound and never a metric of the pipeline. S1
is not deployable either, since the shift presumes onset at the first on-treatment
visit, and it is reported as a diagnostic of where the error sits. The panel~B
labels are the identifiers used in the released code, and the family varies the
combination rule, the direction of the consensus gate and the within-window
aggregation order, all as functions of the same two extracted variables and
therefore at no additional inference cost. C1 and C2b, and C2 and C2\_strictB,
return identical predictions on all 183 cells.}
\label{tab:prompt_variants}
\begin{tabular}{lp{4.6cm}ccccc}
\toprule
\textbf{Variant} & \textbf{Change from the deployed prompt} &
\textbf{Acc.\ (\%)} & \textbf{$\Delta$} & \textbf{Prec.} & \textbf{Rec.} & \textbf{$p_{\text{Holm}}$} \\
\midrule
\multicolumn{7}{l}{\textbf{A. Prompt wording and model count}} \\
\midrule
\textbf{V0} & \emph{Uncontrolled diarrhea.} Base case, as deployed. Both agents exclude adequately controlled diarrhea. & 79.6 $\pm$ 0.3 & --- & 0.596 & 0.660 & --- \\[2pt]
V1 & \emph{Any diarrhea.} Controlled diarrhea reclassified as present. ``Being treated means it is present.'' & 75.2 $\pm$ 0.3 & $-4.4$ & 0.514 & 0.787 & 0.35 \\[2pt]
\textbf{V2} & \emph{Severe diarrhea.} Same exclusion, restated explicitly as a severity bar. ``YES only if active AND clinically significant.'' & \textbf{84.2 $\pm$ 0.0} & $+4.6$ & 0.714 & 0.638 & 0.15 \\[2pt]
V3 & \emph{Severe diarrhea with rules.} V2 plus two reasoning rules. A later resolution governs an earlier onset in the same note, and a finding attributed to another cause is absent. & 83.1 $\pm$ 0.5 & $+3.5$ & 0.700 & 0.596 & 0.35 \\[2pt]
V4 & \emph{Any diarrhea with rules.} V1 plus the same two reasoning rules as V3. & 75.0 $\pm$ 0.3 & $-4.6$ & 0.507 & 0.766 & 0.25 \\[2pt]
Agent~1 & V0 prompt, clinical NLP expert alone, no gate. & 78.5 $\pm$ 0.3 & $-1.1$ & 0.569 & 0.702 & 0.15$^{*}$ \\[2pt]
Agent~2 & V0 prompt, domain specialist alone, no gate. & 79.2 $\pm$ 0.5 & $-0.4$ & 0.593 & 0.681 & 0.19$^{*}$ \\
\midrule
\multicolumn{7}{l}{\textbf{B. Onset/control decomposition, ten combination rules}} \\
\midrule
A\_only     & Onset variable alone, with no control test applied. & 61.7 $\pm$ 0.0 & $-17.9$ & 0.384 & 0.809 & --- \\[2pt]
notB\_only  & Control variable alone, applied as a negation. & 74.7 $\pm$ 0.3 & $-4.9$ & 0.500 & 0.447 & --- \\[2pt]
C1          & First state rule over the two variables. & 75.8 $\pm$ 0.3 & $-3.8$ & 0.526 & 0.426 & --- \\
C2b         & Variant of the second state rule. & 75.8 $\pm$ 0.3 & $-3.8$ & 0.526 & 0.426 & --- \\
C1\_OR      & C1 with the permissive gate on both variables. & 77.0 $\pm$ 0.5 & $-2.6$ & 0.600 & 0.383 & --- \\
C2          & Second state rule over the two variables. & 77.8 $\pm$ 0.3 & $-1.8$ & 0.550 & 0.702 & --- \\
C2\_strictB & C2 with a stricter control test. & 77.8 $\pm$ 0.3 & $-1.8$ & 0.550 & 0.702 & --- \\
C2\_first   & C2 with first-note within-window aggregation. & 78.7 $\pm$ 0.0 & $-0.9$ & 0.567 & 0.723 & --- \\
C2\_OR      & C2 with the permissive gate on both variables. & 79.2 $\pm$ 0.5 & $-0.4$ & 0.589 & 0.702 & --- \\
C2\_any     & C2 with any-note within-window aggregation. & 79.4 $\pm$ 0.3 & $-0.2$ & 0.585 & 0.660 & --- \\
\midrule
\multicolumn{7}{l}{\textbf{C. Temporal alignment of the V0 predictions}} \\
\midrule
S1 & Model output only is slid so that the first visit predicted positive falls at the first on-treatment visit, padding the tail negative. & 84.2 & $+4.4$ & 0.673 & 0.745 & 1.00$^{*}$ \\[2pt]
S3 & Both the predictions and the annotation are aligned to their own first positive visit. Positional, not deployable. & 85.2 & $+5.5$ & 0.692 & 0.766 & --- \\
\bottomrule
\multicolumn{7}{l}{\small $^{*}$Contrast is against V2 rather than against V0.} \\
\end{tabular}
\end{table}

\begin{table}[H]
\centering
\small
\caption{Architecture selection, reported on the numbers it was decided with
(Note~S2). Six methods of increasing sophistication were evaluated on the identical
189-cell diarrhea ground truth under the original note corpus and the order-date
exposure anchor, and this comparison is what selected the two-agent consensus
architecture that the reported pipeline holds fixed. These are \emph{not} the
reported metrics, and they are not comparable to main-text Table~2 or to
Table~\ref{tab:prompt_variants}, both of which run on the refreshed corpus, the
corrected anchor and a 183-cell grid. Panel~A gives accuracy as the mean$\pm$SD over
three runs for the LLM methods, with the three deterministic methods exact, and the
best value in each column in bold. Panel~B gives all 15 pairwise McNemar exact
tests, Holm-adjusted across the 15 comparisons, where $b$ counts the cells on which
the first method is correct and the second wrong. Only the contrasts against the
regex floor are significant, so the architecture was adopted as the best-performing
configuration rather than as a proven necessity. The final rerun confirms that
reading by placing the gate within 0.4~points of the stronger single agent under the
deployed prompt, differing on two of 183~cells (Note~S1). The consensus accuracy here,
76.4\%, is not the 77.1\% that Note~S8 reports for the deployed prompt on a
189-cell grid, because this comparison is an earlier stage of development that
ran on a different note pull and a different set of evaluation windows.}
\label{tab:architecture}
\begin{tabular}{llcccc}
\toprule
\multicolumn{6}{l}{\textbf{A. Six methods on the 189-cell grid}} \\
\midrule
\multicolumn{2}{l}{\textbf{Method}} & \textbf{Accuracy (\%)} & \textbf{Precision} & \textbf{Recall} & \textbf{F1} \\
\midrule
\multicolumn{2}{l}{Regex-only (floor)}     & 42.3 & 0.293 & \textbf{0.936} & 0.447 \\
\multicolumn{2}{l}{Rule-based NLP}         & 70.9 & 0.423 & 0.468 & 0.444 \\
\multicolumn{2}{l}{Hybrid NLP}             & 70.9 & 0.436 & 0.575 & 0.495 \\
\multicolumn{2}{l}{Single-LLM}             & 72.7 $\pm$ 0.6 & 0.458 $\pm$ 0.010 & 0.532 $\pm$ 0.000 & 0.492 $\pm$ 0.006 \\
\multicolumn{2}{l}{Dual-LLM (permissive)}  & 73.0 $\pm$ 0.5 & 0.464 $\pm$ 0.008 & 0.553 $\pm$ 0.000 & 0.505 $\pm$ 0.005 \\
\multicolumn{2}{l}{\textbf{Dual-LLM consensus}} & \textbf{76.4 $\pm$ 0.3} & \textbf{0.527 $\pm$ 0.007} & 0.489 $\pm$ 0.000 & \textbf{0.507 $\pm$ 0.003} \\
\midrule
\multicolumn{6}{l}{\textbf{B. All 15 pairwise McNemar exact tests}} \\
\midrule
\textbf{Method A} & \textbf{Method B} & \textbf{$b$} & \textbf{$c$} & \textbf{$p$} & \textbf{$p_{\text{Holm}}$} \\
\midrule
Regex-only     & Rule-based NLP & 22 & 76 & \textbf{$<$0.0001} & \textbf{$<$0.0001} \\
Regex-only     & Hybrid NLP     & 18 & 72 & \textbf{$<$0.0001} & \textbf{$<$0.0001} \\
Regex-only     & Single-LLM     & 19 & 77 & \textbf{$<$0.0001} & \textbf{$<$0.0001} \\
Regex-only     & Permissive     & 18 & 77 & \textbf{$<$0.0001} & \textbf{$<$0.0001} \\
Regex-only     & Consensus      & 21 & 85 & \textbf{$<$0.0001} & \textbf{$<$0.0001} \\
\addlinespace
Rule-based NLP & Hybrid NLP     &  7 &  7 & 1.000 & 1.000 \\
Rule-based NLP & Single-LLM     & 13 & 17 & 0.585 & 1.000 \\
Rule-based NLP & Permissive     & 12 & 17 & 0.458 & 1.000 \\
Rule-based NLP & Consensus      & 11 & 21 & 0.110 & 1.000 \\
Hybrid NLP     & Single-LLM     & 15 & 19 & 0.608 & 1.000 \\
Hybrid NLP     & Permissive     & 14 & 19 & 0.487 & 1.000 \\
Hybrid NLP     & Consensus      & 14 & 24 & 0.143 & 1.000 \\
\addlinespace
Single-LLM     & Permissive     &  0 &  1 & 1.000 & 1.000 \\
Single-LLM     & Consensus      &  2 &  8 & 0.109 & 1.000 \\
Permissive     & Consensus      &  3 &  8 & 0.227 & 1.000 \\
\bottomrule
\end{tabular}
\end{table}

\clearpage
\section*{Supplementary Figures}

\begin{suppfigure}[H]
\centering
\includegraphics[width=\textwidth]{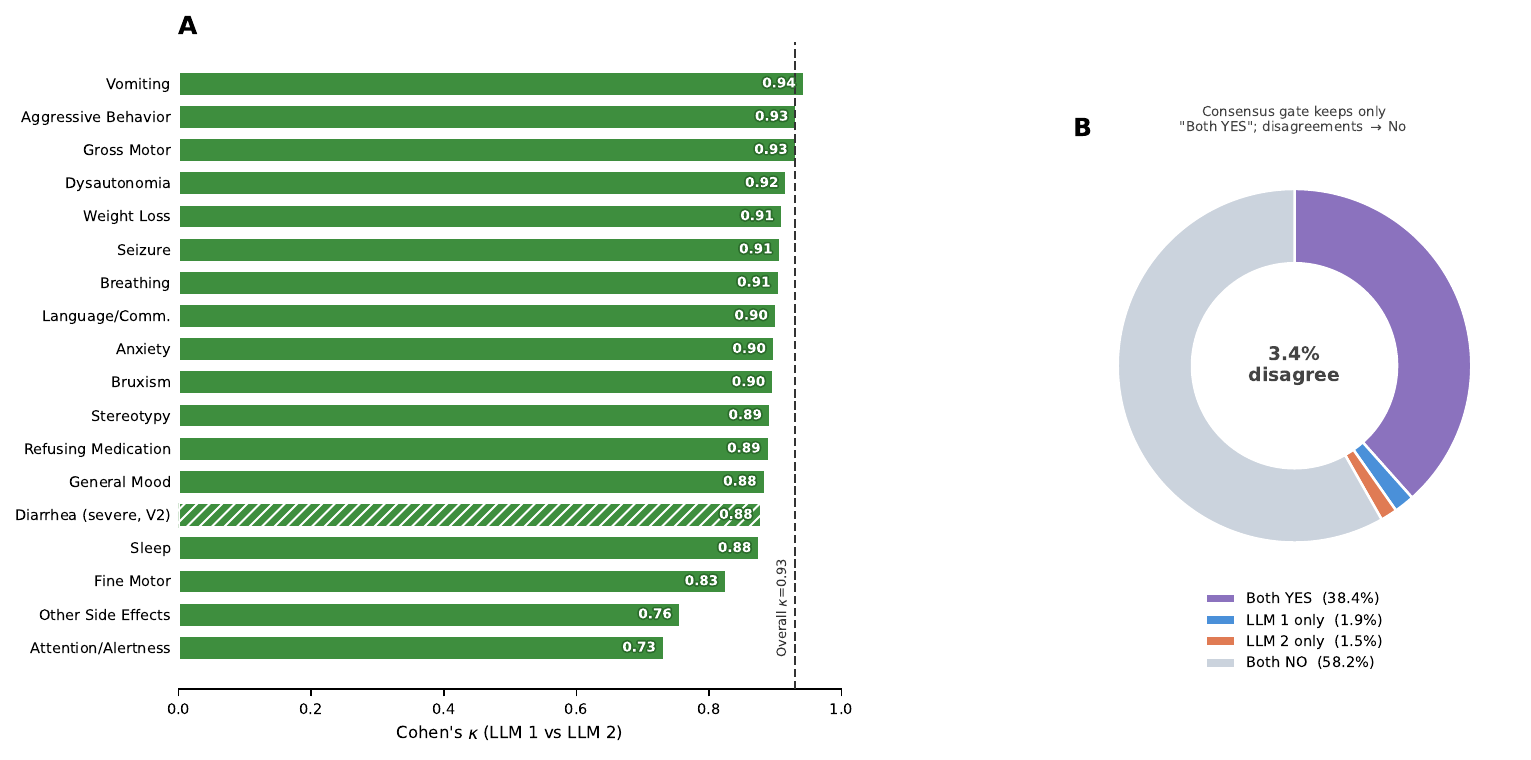}
\caption{Inter-LLM agreement supports the consensus-gate design
(Claude~Opus~4.6). \textbf{(A)}~Cohen's $\kappa$ between the two agents for each
phenotype, sorted descending. Every phenotype reaches the substantial band, so the
bars share one color, and the dashed line is the overall $\kappa=0.93$ across the
60,530 instances both agents scored. Agreement is highest for discrete, observable
events such as vomiting, aggressive behavior and gross-motor findings, and lowest
for attention/alertness, the phenotype with the fewest instances (313), and for the
open-ended other-side-effects category. The hatched bar is diarrhea, scored by the
diarrhea-specific prompt on its own validated corpus rather than by the presence
prompt, and shown for reference. \textbf{(B)}~Consensus outcome composition over the
same 60,530 matches. The gate retains only ``both YES'' (38.4\%) as positive, and
the 3.4\% one-sided disagreements are conservatively classified negative.}
\label{fig:agreement}
\end{suppfigure}

\clearpage
\bibliography{references}